\documentclass[prb,amsmath,amssymb,superscriptaddress,twocolumn]{revtex4}

\usepackage{amsmath}
\usepackage{graphicx}
\usepackage{multirow}
\usepackage{subfig}
\usepackage{bbm,color,ulem}
\usepackage{dsfont}
\usepackage{float}
\allowdisplaybreaks

\DeclareMathOperator{\sgn}{sgn}

\DeclareMathOperator{\re}{Re}
\DeclareMathOperator{\im}{Im}

\begin{document}
\title{Coupled electron-impurity and electron-phonon systems as trivial non-Fermi liquids}
\author{Donovan Buterakos}
\author{Sankar Das Sarma}
\affiliation{Condensed Matter Theory Center and Joint Quantum Institute, Department of Physics, University of Maryland, College Park, Maryland 20742-4111 USA}
\date{\today}

\begin{abstract}
We consider an electron gas, both in two (2D) and three (3D) dimensions, interacting with quenched impurities and phonons within leading order finite-temperature many body perturbation theories, calculating the electron self-energies, spectral functions, and momentum distribution functions at finite temperatures.   The resultant spectral function is in general highly non-Lorentzian, indicating that the system is not a Fermi liquid in the usual sense.  The calculated momentum distribution function cannot be approximated by a Fermi function at any temperature, providing a rather simple example of a non-Fermi liquid with well-understood properties.
\end{abstract}

\maketitle

\section{Introduction}

The concept of a Fermi liquid is central to our understanding of the properties of metals and He-3.  The idea, as propounded by Landau more than 60 years ago, is that an interacting Fermi liquid is adiabatically connected to the non-interacting Fermi gas, thus preserving a one-to-one correspondence between interacting and non-interacting distribution functions.  In particular, the concept of a Fermi surface is preserved in the interacting system albeit with many-body renormalizations of various single-particle parameters such as the effective mass and the Lande g-factor.  The interacting spectral function thus must have a delta function implying the existence of well-defined quasiparticles with the imaginary part of the self-energy vanishing as $E^2$ (or $T^2$) close to the Fermi surface.  This also immediately implies that the interacting zero-temperature momentum distribution function has a discontinuity at the Fermi momentum although the size of this discontinuity is suppressed from its non-interacting value due to many-body renormalization by electron-electron interactions. This is certainly true for electron-electron interactions in 3D and 2D (there are logarithmic corrections in 2D to the self-energy which do not affect these considerations), and it is well-known that interacting 2D and 3D electron systems are indeed Fermi liquids except perhaps at exponentially low temperatures of no physical interest where non-perturbative effects could give rise to superconductivity, the so-called Kohn-Luttinger superconductivity, at high orbital momentum channels.\cite{Abrikosov1963, Nozieres1964, ShankarRMP1994}

Much interest has focused on situations where such a Landau Fermi liquid paradigm does not apply, and interactions lead to a breakdown of the Fermi liquid picture destroying the one-to-one correspondence between noninteracting and interacting systems with the disappearance of the Fermi surface in the interacting system.  Such systems, where the Fermi liquid paradigm fails, are called ``non-Fermi liquids'' (NFL).  The most well-known example of such an NFL is the interacting 1D electron system, called the Luttinger liquid, where interactions lead to the destruction of the Fermi surface and the one-to-one correspondence between interacting and noninteracting systems.\cite{Giamarchi2003, HaldaneJPC1981, VoitRPP1995}  In a Luttinger liquid, the zero-temperature momentum distribution function is continuous through the Fermi momentum, characterized by an interaction-dependent exponent called the Luttinger exponent.  There is a widespread belief that many strongly correlated materials are also NFLs, possibly arising from the existence of interaction-driven quantum critical transitions in the ground state.  In particular, high-temperature superconducting cuprates are often thought to be NFLs in their normal state above $T_c$ with the terminology ``strange'' (or ``bad'') metals, in contrast to normal FL metals, applied to emphasize the possibly NFL nature of the system.  In spite of very extensive research on the subject, it is rather difficult to establish the NFL nature of a system starting from a realistic interacting Hamiltonian, and often, theories assume the existence of an underlying NFL on phenomenological grounds.  This is in sharp contrast to 1D interacting systems and normal 3D metals, where the existence of an NFL, namely the Luttinger liquid, and of an FL, respectively, is definitively established.  Finding generic 2D or 3D NFLs starting from reasonable (and not fine-tuned or phenomenological) microscopic models has been one of the most important open problems in many body condensed matter theory for a long time.

The goal of the current work is to ask what happens in electron systems when electron-electron interaction effects are ignored, but other interactions invariably present in real materials, such as electron-impurity interactions or electron-phonon interactions are included in calculating the properties of the electron system.  We are therefore asking a question complementary to that discussed above for NFLs: does electron-impurity or electron-phonon interaction, by itself, lead to a Fermi liquid in the strict sense of a one-to-one correspondence between interacting and noninteracting systems?

The answer, although simple, turns out to be somewhat surprising and not widely mentioned in many-body textbooks.  Electron-impurity and electron-phonon interactions invariably lead to violations of the Landau paradigm, and a coupled electron-impurity or coupled electron-phonon system, without any electron-electron interaction, is in fact a trivial NFL!  Thus, the existence of NFLs in real materials is generic since all systems have impurities and phonons, and electrons interacting with impurities and/or phonons are generically non-Fermi liquids albeit in very different manners.  An intuitive way of understanding this is that impurity scattering leads to a finite imaginary part of the self-energy even on the Fermi surface in contrast to electron-electron interaction where the imaginary self-energy vanishes as $E^2$, with $E$ being the energy deviation from the Fermi surface.  A finite imaginary self-energy immediately implies that the spectral function of the coupled electron-impurity system cannot be a strict delta function, and hence the momentum distribution function is smooth with no discontinuity even at $T=0$.  For very strong impurity scattering, a situation not considered in our work, the electrons become Anderson-localized, and such a localized system is obviously a trivial NFL with no one to one correspondence with the non-interacting Fermi gas.  The coupled electron-phonon system must have spectral features at the phonon energy arising from electron-phonon interactions, rendering the electron spectral function strongly non-Lorentzian with no strict  delta function like features necessary for obtaining the usual FL quasiparticles.  But the distribution function for the coupled electron-phonon system still has a finite discontinuity (at $T=0$) with the size of the discontinuity determined by the electron-phonon coupling. Depending on the details of the electron-phonon coupling, it is possible that the electronic spectral function manifests nonquasiparticle-type behavior down to rather low temperatures in the presence of electron-phonon interaction (although in normal 3D metals, the effects of phonon coupling are strongly quenched for $T<50K$, which is still an extremely low temperature on the electronic scale since the typical metallic Fermi temperature is $\sim$ 50,000K).

We are, of course, not the first to investigate the quasiparticle nature of coupled electron-phonon or electron-impurity systems.  The issue has been studied extensively going back 60 years, and the spectral function of interacting electron-phonon systems has been studied in various realistic models extensively.\cite{Grimvall1981} Kadanoff,\cite{KadanoffPR1963} and then Prange and Kadanoff,\cite{PrangePR1964} were among the first to point out the non-quasiparticle like behavior of the spectral function in the coupled electron-phonon system, with Ref. \onlinecite{KadanoffPR1963} having the striking title: ``Failure of the electronic quasiparticle picture for nuclear spins relaxation in metals''.  In modern terminology, the failure of the quasiparticle picture is the primary characteristic of an NFL.  Reference \onlinecite{PrangePR1964}, in developing a transport theory for electron-phonon interactions in metals, emphasizes in its abstract: ``the electronic excitation spectrum has considerable width and structure so that one might not expect a priori that there would be well-defined quasiparticles.''

We study three separate finite temperature many-body problems theoretically in this paper: A coupled electron-phonon system in 2D and 3D using (1) Einstein, and (2) Debye phonon models, and (3) a coupled electron-impurity problem (in 2D and 3D as well as in 1D) using a model impurity scattering potential.  We obtain the  electron self-energy, the interacting spectral function, and the momentum distribution function in each case, giving results both as functions of energy and temperature.  We discuss in some depth the relationship between the interacting distribution function and the non-interacting Fermi distribution function, critically commenting on the extent to which the system supports Landau type quasiparticles and can be construed to be an NFL.  We consider only the leading-order diagrams in the electron-phonon or electron-impurity interaction perturbatively, but our qualitative conclusion should apply for stronger interactions.  Our calculations are essentially analytical, thus shedding considerable light on the nature of rather simple NFLs which may arise from interactions with phonons or impurities outside the Landau FL paradigm. We emphasize that we neglect all effects of electron-electron interactions in the theory (except perhaps indirectly in determining the electron-phonon and/or electron-impurity coupling, treated as parameters in our theory, through screening effects), and thus, our work has no relevance to the important issue of the possible existence or not of NFLs in strongly correlated materials where the question is whether electron-electron interactions can possibly drive the system into an NFL.  Our purpose is to demonstrate through concrete calculations that apparent nonquasiparticle-like behavior could arise trivially from finite temperature electron-phonon or electron-impurity interactions.

The rest of the paper is organized as follows.  In sec. II we study the electron-phonon interaction using the Einstein phonon model; in sec. III we study the electron-phonon interaction using the Debye phonon model; and in sec. IV, we study the electron-impurity interaction problem.  We conclude in sec. V providing a discussion.

\section{Electron-Phonon Interactions in the Einstein Model}

We investigate an electron gas with electron-phonon interactions. The general Hamiltonian of this system is given by:
\begin{align}
H_{\text{el-ph}}&=\sum_{k}E_kc_k^\dagger c_k+\sum_{q}\omega_{\text{ph}}(q)\,a_q^\dagger a_q\nonumber\\&+\sum_{k,k'}g_{k,k'}\,c_k^\dagger c_{k'}(a_{k'-k}^\dagger+a_{k-k'})
\label{eqn:elphh}
\end{align}

We first consider the Einstein model for phonons; that is, we assume that the phonon dispersion relation is constant, $\omega_{\text{ph}}(q)=\omega_0$. We also assume that the the electron-phonon coupling constant is momentum-independent, $g_{k,k'}=g$. Using many-body Matsubara techniques, we calculate the finite-temperature electron self-energy to leading order in $g^2$. With this result, we then obtain and plot the spectral function and momentum distribution function, which manifest apparent NFL behavior at finite temperatures.

\subsection{Self-Energy Calculation}

\begin{figure}[!htb]
	\centering
	\includegraphics[width=.7\columnwidth]{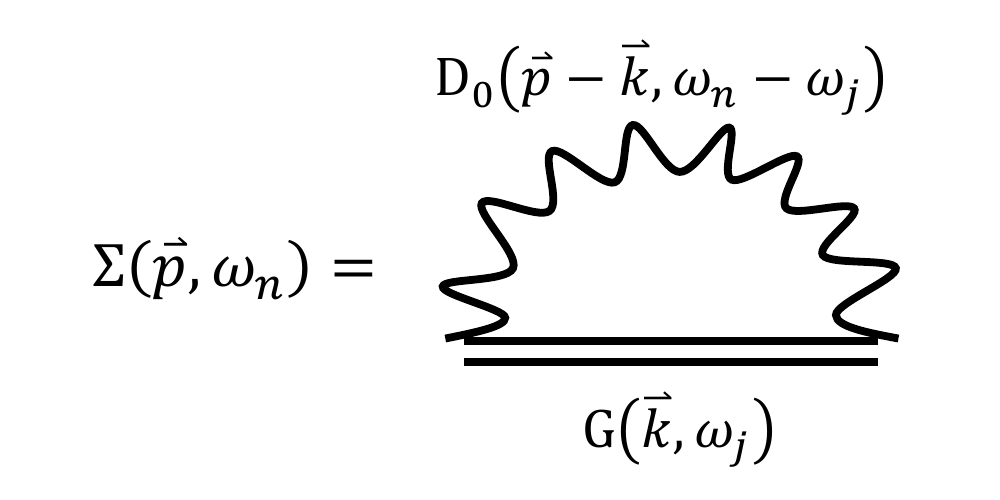}
	\caption{The diagram for the electron's self-energy. In our calculation, the electron Green's function $\mathcal{G}$ includes the electron's self energy $\Sigma$; however the bare phonon Green's function $\mathcal{D}_0$ is used.}
	\label{fig:diagram}
\end{figure}

We perform a similar calculation to the one initially done in Ref. \onlinecite{EnglesburgPR1963}, except we consider the case of finite temperature. To first order, the self energy is given by the Fock diagram pictured in fig. \ref{fig:diagram}. In this calculation, we use the bare phonon Green's function $\mathcal{D}_0(\vec{q},\omega_i)$; however, we attempt to enforce a self-consistency condition by using the dressed electron Green's function $\mathcal{G}(\vec{k},\omega_j)$, which is dependent on the electron's self-energy. Using thermal Feynman rules to evaluate this diagram, we obtain the following expression in $d$ dimensions:
\begin{align}
	\Sigma(\vec{p},\omega_n)=-g^2T\sum_{\omega_j}\int\frac{d^dk}{(2\pi)^d}&\frac{-\omega_0^2}{(\omega_n-\omega_j)^2+\omega_0^2}\times\nonumber\\&\frac{1}{i\omega_j-E_k-\Sigma(\vec{k},\omega_j)}
	\label{eqn:initialsigma}
\end{align}

where the sum is over Matsubara frequency $\omega_j=j\pi T$ for odd integer $j$, and $E_k$ is the energy of an electron with momentum $k$ measured from the chemical potential $\mu$. From this expression, we note that $\Sigma(\vec{p},\omega_n)$ is independent of $\vec{p}$, and thus we can remove the $\vec{k}$-dependence of $\Sigma$ inside the integral. If we approximate the density of states as being $k$-independent $N_d(E_k)\approx N_d(\mu)$, we can rewrite the $k$ integral as an integral over $E_k$, which we then evaluate by extending the lower limit of integration to negative infinity.
\begin{align}
&\int\frac{d^dk}{(2\pi)^d}\,\frac{1}{i\omega_j-E_k-\Sigma(\omega_j)}\nonumber\\
\approx&\,N_d\int_{-\infty}^\infty dE\,\frac{1}{i\omega_j-E-\Sigma(\omega_j)}=-i\pi N_d\sgn\omega_j
\end{align}

Because of the $\sgn\omega_j$ term, all terms in the sum with $\omega_j<0$ cancel those with 
$\omega_j>2\omega_n$, as follows:
\begin{align}
	\Sigma(\omega_n)&=i\pi g^2TN_d\sum_{\omega_j}\frac{-\omega_0^2}{(\omega_n-\omega_j)^2+\omega_0^2}\sgn\omega_j\nonumber\\
	&=-i\pi g^2TN_d\sgn\omega_n\sum_{\omega_l=-\omega_{|n|-1}}^{\omega_{|n|-1}}\frac{\omega_0^2}{\omega_l^2+\omega_0^2}
\end{align}

where the sum is over all $\omega_l=l\pi T$ for even integer $l$ such that $|\omega_l|<|\omega_n|$. This sum can be written in closed form using the digamma function $\psi^{(0)}(z)$, and since the digamma function is analytic, it is trivial to analytically continue the result to the upper half of the complex plane. Thus the retarded self-energy is obtained as:
\begin{align}
	&\Sigma_R(\omega)=\frac{g^2N_d\omega_0}{2}\Bigg[-\pi i\coth\frac{\omega_0}{2T}\nonumber\\&+\psi^{(0)}\Big(\frac{1}{2}+i\frac{\omega_0-\omega}{2\pi T}\Big)-\psi^{(0)}\Big(\frac{1}{2}+i\frac{-\omega_0-\omega}{2\pi T}\Big)\Bigg]
	\label{eqn:srein}
\end{align}

We note that Eq. (\ref{eqn:srein}) for the electron self-energy is dimension-independent in our model, following Ref. \onlinecite{EnglesburgPR1963}, i.e., all information about spatial dimensions is hidden in the density of states $N_d$ which we take to be a model parameter.  In addition, the self-energy is also momentum independent by virtue of the momentum independence of the electron-phonon coupling in our model, which is parametrized by the coupling strength $g$.

\begin{figure}[!htb]
	\centering
	\includegraphics[width=\columnwidth]{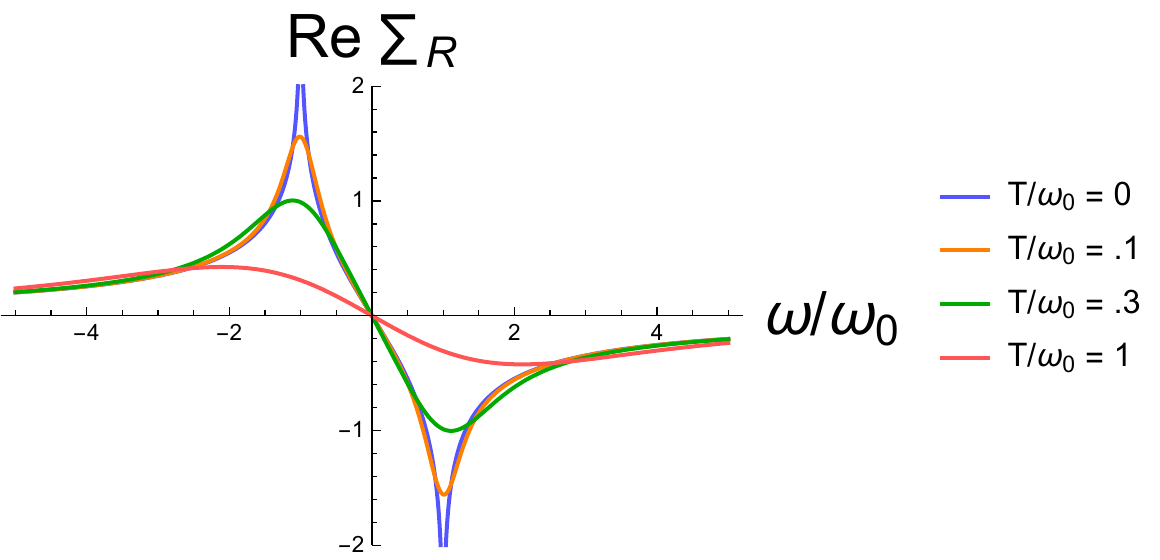}	
	\includegraphics[width=\columnwidth]{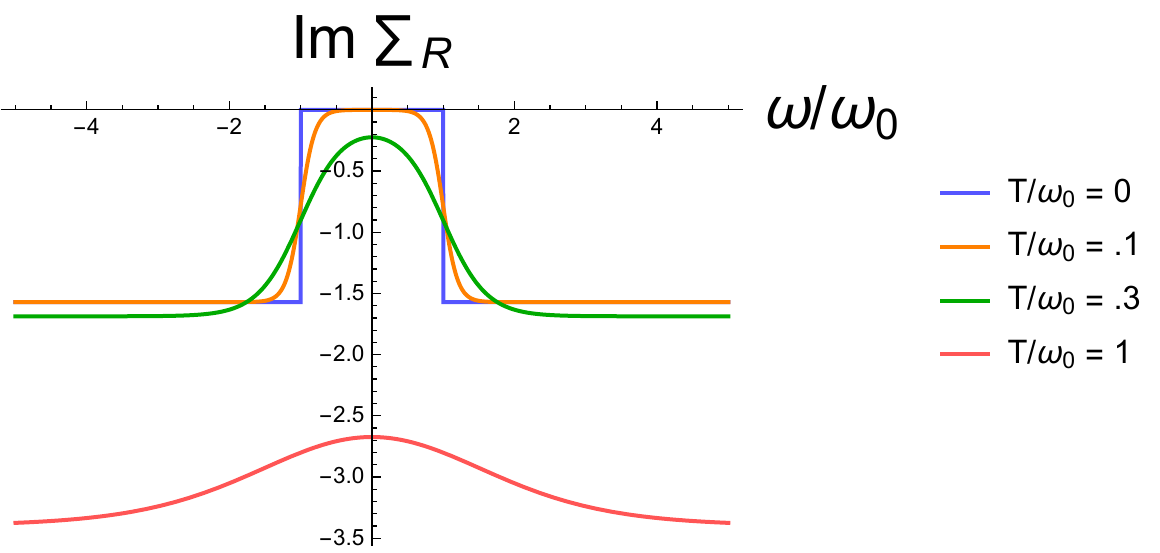}
	\caption{Plot of $\Sigma_R/(\omega_0 g^2N_d)$ versus $\omega/\omega_0$ with $T$ held constant for the Einstein phonon model.}
	\label{fig:esigma}
\end{figure}

\begin{figure}[!htb]
	\centering
	\includegraphics[width=\columnwidth]{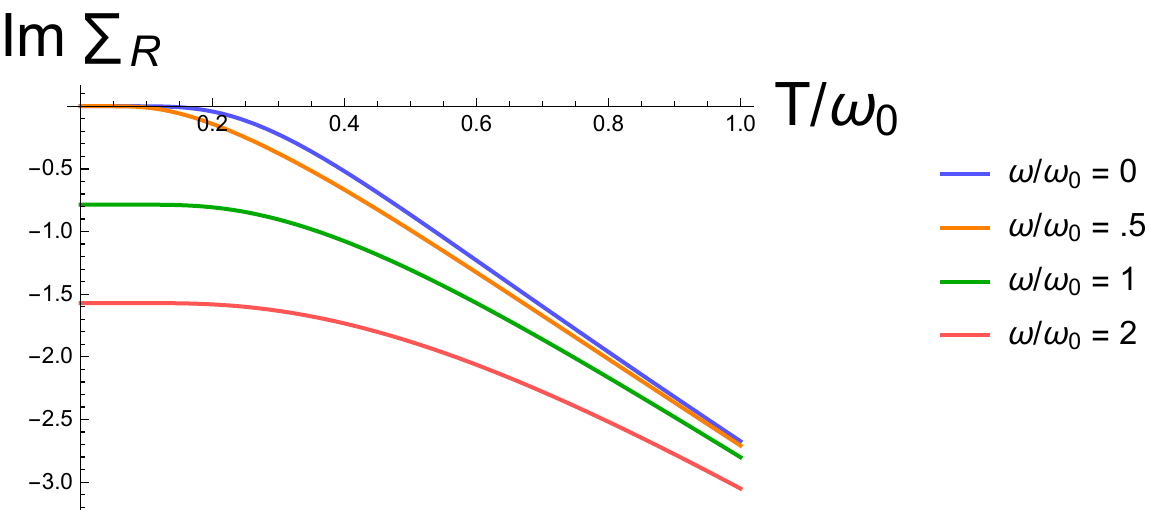}
	\caption{Plot of $\im\Sigma_R/(\omega_0 g^2N_d)$ versus $T/\omega_0$ with $\omega$ held constant for the Einstein phonon model.}
	\label{fig:esigmat}
\end{figure}

We show the calculated self-energy in figs. \ref{fig:esigma} and \ref{fig:esigmat}, respectively as functions of $\omega$ and $T$.  An important thing to note is that the imaginary part of the self-energy vanishes for small $\omega$ and $T$ guaranteeing the FL behavior as $T/\omega_0$ vanishes, but since the energy scale is set by the phonon energy $\omega_0$, the system could manifest an apparent NFL behavior down to low temperatures if $\omega_0$ happens to be small.  Thus, when the phonons are soft (i.e. low $\omega_0$), the coupled electron-phonon system is an apparent NFL down to low $T$ with $\im\Sigma_R\sim T$.  One direct experimental implication of this result is that for systems with low-energy phonon modes, the linear-in-$T$ resistivity (corresponding to the linear regime in $\im\Sigma_R$ for $T/\omega_0>0.2$ in Fig. \ref{fig:esigmat}) could persist to rather low temperatures as emphasized recently.\cite{HwangPRB2019}


\subsection{Expansions and Discussion}

\begin{figure*}[!htb]
	\centering
	\includegraphics[width=2\columnwidth]{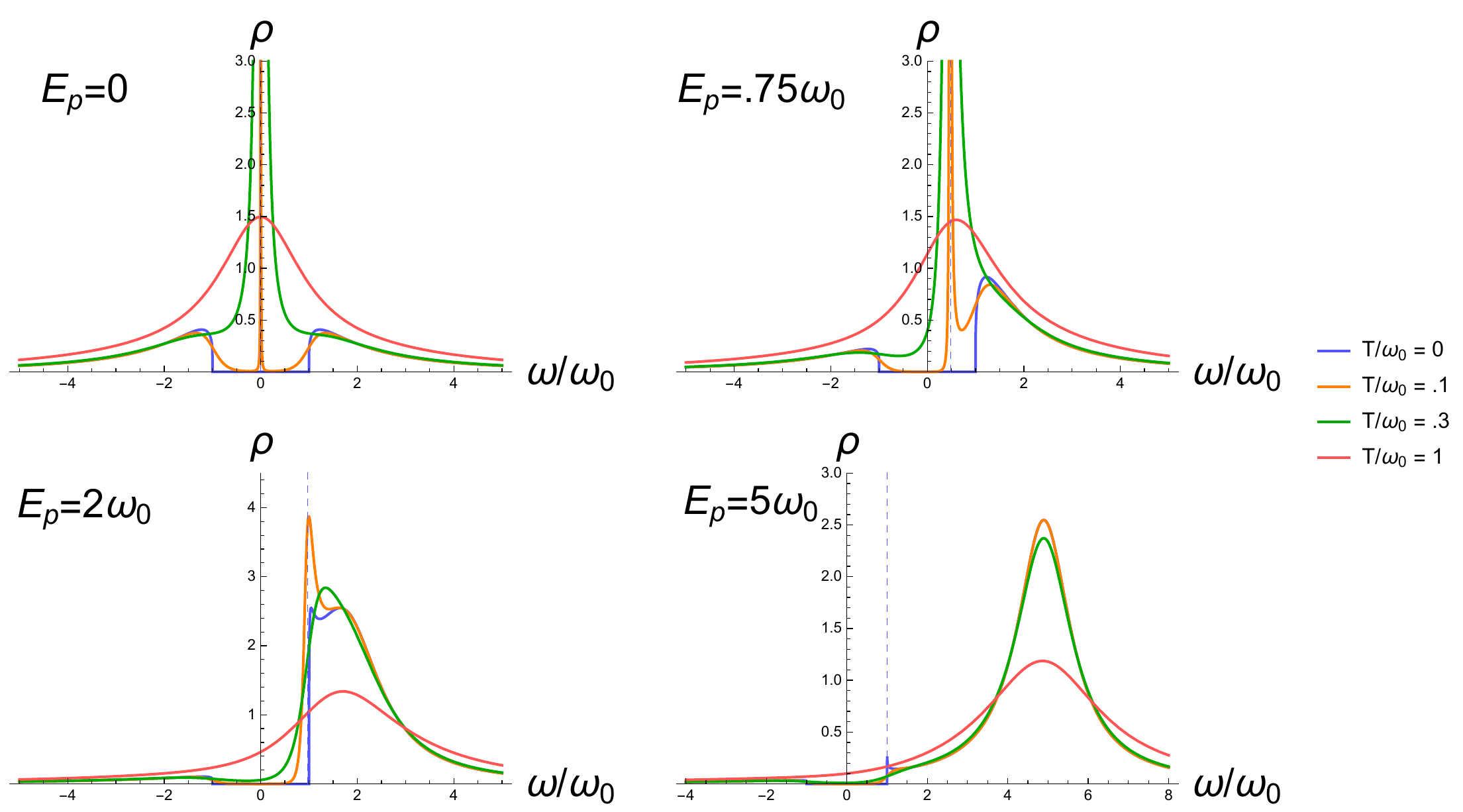}	
	\caption{Plots of $\rho(E_p,\omega)$ with $g^2N=0.5$ for different values of $T$ and $E_p$ for the Einstein phonon model. The dashed line represents a delta function present in the $T=0$ case.}
	\label{fig:erho}
\end{figure*}

Using the reflection formula for the digamma function, one can write $\im\Sigma_R$ in terms of elementary functions:
\begin{align}
&\im\Sigma_R(\omega)=\nonumber\\&\frac{\pi g^2N_d\omega_0}{4}\bigg(\tanh\frac{\omega_0-\omega}{2T}+\tanh\frac{\omega_0+\omega}{2T}-2\coth\frac{\omega_0}{2T}\bigg)
\end{align}

We expand eq. (\ref{eqn:srein}) for $T\ll\omega_0,\omega_0\pm\omega$ to obtain the low-temperature behavior of the self-energy, as follows:
\begin{align}
\Sigma_R(\omega)&=\frac{g^2N_d\omega_0}{2}\Bigg[-\pi i\,\Theta(\omega^2-\omega_0^2)+\log\bigg|\frac{\omega-\omega_0}{\omega+\omega_0}\bigg|\nonumber\\&+\frac{\pi^2}{6}\bigg(\frac{1}{(\omega+\omega_0)^2}-\frac{1}{(\omega-\omega_0)^2}\bigg)T^2+O\big(T^3\big)\Bigg]
\end{align}

where $\Theta(x)$ is the Heaviside step function. Similarly, we expand eq. (\ref{eqn:srein}) for $T\gg\omega_0,\omega$ to obtain the high-temperature behavior:

\begin{align}
\Sigma_R(\omega)=\frac{g^2N_d\omega_0}{2}\Bigg[&\,i\,\bigg(\frac{-2\pi T}{\omega_0}+\frac{\pi\omega_0}{3T}\bigg)\nonumber\\&+\frac{\psi^{(2)}(\frac{1}{2})\omega\omega_0}{2\pi^2T^2}+O\bigg(\frac{1}{T^3}\bigg)\Bigg]
\label{eqn:esigmalarget}
\end{align}

We emphasize that the linear-in-$T$ behavior of $\im\Sigma_R$ in eq. (\ref{eqn:esigmalarget}) is generic for $T>\omega_0$, and is a so-called ``marginal Fermi liquid'' behavior which in this case arises from coupling to phonons, and could persist to rather low temperatures ($T\sim 0.2\omega_0$).  Experimentally, there is no way to distinguish such a marginal Fermi liquid behavior arising from phonons from that arising from electron-electron interactions except to go to $T<0.2 \omega_0$ which may not be feasible in specific experimental situations if the phonons are soft.  We mention that the same behavior will manifest itself if phonons in the current model are replaced by some other bosonic modes (e.g. magnons, Goldstone modes associated with a quantum critical point) as long as the same model as what we use here is valid for the relevant electron-boson coupling. We note that the calculation we performed is valid in all dimensions, with the only difference being in the density of states $N_d$.

Using the results in eq. (\ref{eqn:srein}), we plot the calculated electron spectral function $\rho(E_p,\omega)=-2\im G^R(E_p,\omega)$ in fig. \ref{fig:erho}, setting $g^2N_d=1/2$, and $E_p/\omega_0=\{0,\,.75,\,2,\,5\}$ to maintain consistency with Ref. \onlinecite{EnglesburgPR1963} where the corresponding $T=0$ theory was developed. We see in fig. \ref{fig:erho} that the low temperature curves ($T<\omega_0$) very closely approximate the zero-temperature results, but at a scale of around $T\sim\omega_0$, the spectral function begins to lose its distinctive quasiparticle features, approaching a simple incoherent Lorentzian curve. Therefore, for low $\omega_0$, the spectral function is mostly of an incoherent NFL-like form although it recovers the FL form for $T\ll\omega_0$.

\begin{figure}[!htb]
	\centering
	\includegraphics[width=\columnwidth]{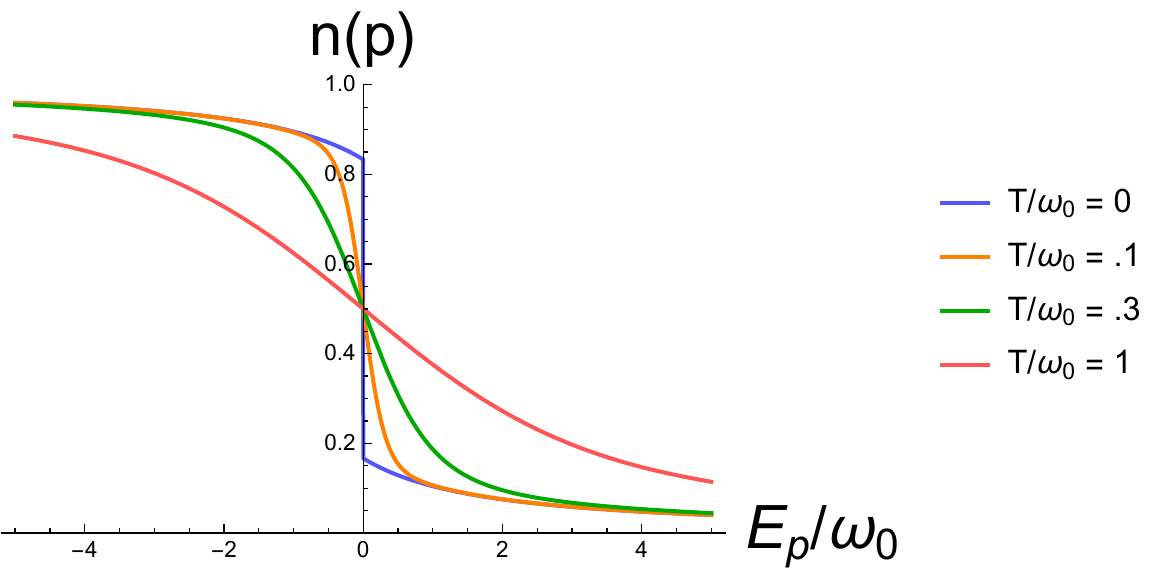}	
	\caption{Plot of $n(p)$ for different values of $T$ for the Einstein phonon model.}
	\label{fig:en}
\end{figure}

From this, we calculate the momentum distribution function $n(p)$ by numerically integrating $\int\frac{d\omega}{2\pi}\rho(p,\omega)f(\omega)$, where $f(\omega)$ is the Fermi-Dirac distribution. The calculated Fermi distribution function is shown in fig. \ref{fig:en}, and has the well-known FL-form at $T=0$ with a finite discontinuity at the Fermi surface.  But, the finite temperature Fermi distribution function is generically NFL-like, particularly for $T\sim\omega_0$.

%
%
%

\section{Electron-Phonon Interactions in the Debye Model}

\subsection{Self-Energy Calculation}

We now consider the Debye phonon model, where phonon frequency is proportional to its momentum up to a cutoff frequency $\omega_D$, the so-called Debye frequency. The Hamiltonian for this system is again given by eqn. \ref{eqn:elphh} with the phonon dispersion now given by $\omega_{\text{ph}}(q)=cq\,\Theta(q-q_D)$, with $q_D$ being the Debye momentum. Additionally, we use the explicit form of the electron dispersion $E_k=k^2/2m-\mu$. Then, proceeding similarly to the leading-order calculation in the previous section, with $q=|\vec{p}-\vec{k}|$, the self energy is given by:
\begin{align}
\Sigma(\vec{p},\omega_n)=-g^2T\sum_{\omega_j}\int&\frac{d^dk}{(2\pi)^d}\frac{-c^2q^2}{(\omega_n-\omega_j)^2+c^2q^2}\times\nonumber\\&\frac{1}{i\omega_j-E_k-\Sigma(\vec{k},\omega_j)}
\end{align}

where the region of integration is the sphere of radius $q_D$ centered at the point $\vec{p}$. This makes it necessary to change the variables of integration to $k$ and $q$ as follows:
\begin{align}
d^2k&=\frac{2kq(k^2-q^2)}{p^2\sqrt{-(k^2-p^2)^2+2q^2(k^2+p^2)-q^4}}dk\,dq\nonumber\\
d^3k&=\frac{kq(k^2-q^2)}{p^3}dk\,dq\,d\varphi
\label{eqn:varchange}
\end{align}

We note that in the 2D case, an additional factor of 2 must also be included since $(k_x,k_y)\rightarrow(k,q)$ is a 2-to-1 mapping. We proceed by assuming $p\approx k_0$, where $k_0$ is defined to be $\sqrt{2m\mu}$ (at $T=0$, $k_0$ is simply the Fermi momentum, but $k_0$ differs from $k_F$ at finite $T$), and evaluating the integrals to leading order in $q_D/k_0$, again ignoring the $k$-dependence of $\Sigma$ inside the integral. The $k$ integral then takes the form:
\begin{align}
&\int_{p-q}^{p+q}k\,dk\,A(k)\,\frac{1}{i\omega_j-E_k-\Sigma(k,\omega_j)}\nonumber\\
\approx&\int_{-qv_F}^{qv_F}\!m\,dE_k\,A(k)\,\frac{1}{i\omega_j-E_k-\Sigma(\omega_j)}\nonumber\\\approx&-i\pi mA(k_0)\sgn\re\omega_j
\label{eqn:approxeint}
\end{align}




Where $A(k)$ denotes the $k$-dependence of the rest of the integrand (from eq. \ref{eqn:varchange}). In order to evaluate the integral, the limits of integration have been extended to $\pm\infty$. This assumption breaks down for $|p-k_0|\sim q_D$, since, for example, when $p=k_0+q_D$, the lower limit of the integral with respect to $E_k$ becomes positive and thus eq. \eqref{eqn:approxeint} cannot be used. Thus our results are valid only in the regime where $E_p\ll q_D\mu/k_0$. Then we have, for $d=2$ or $3$:
\begin{align}
\Sigma(\omega_n)=\frac{-2g^2Tk_0}{(2\pi)^2\mu}\sum_{\omega_j}\int_0^{q_D}&dq\Big(\frac{q}{2}\Big)^{d-2}(-i\pi\sgn\re\omega_j)\times\nonumber\\&\frac{-c^2q^2}{(\omega_n-\omega_j)^2+c^2q^2}
\end{align}

%

\begin{figure}[!htb]
	\centering
	\includegraphics[width=\columnwidth]{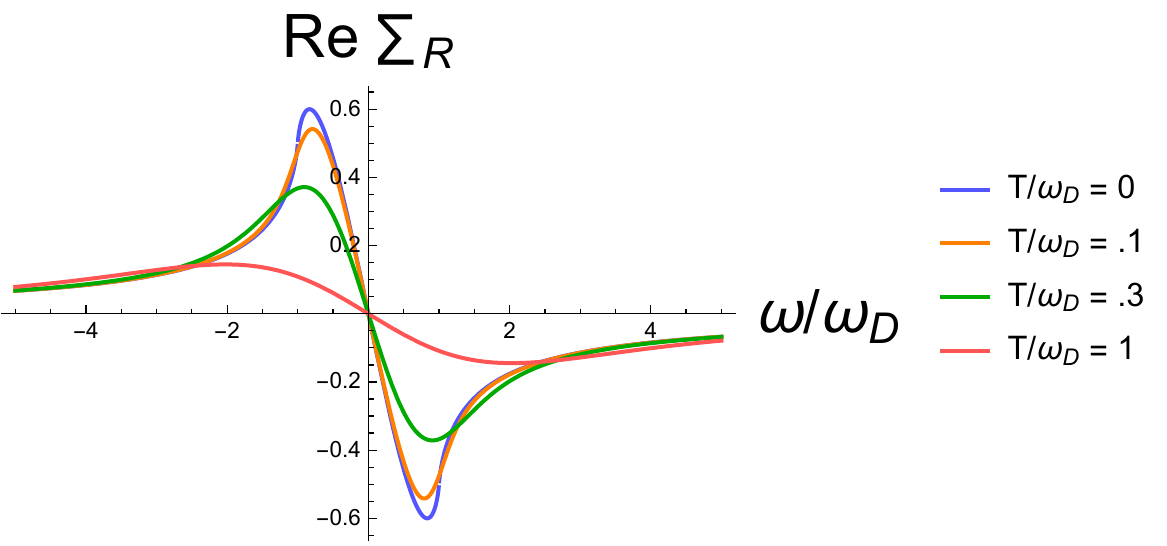}	
	\includegraphics[width=\columnwidth]{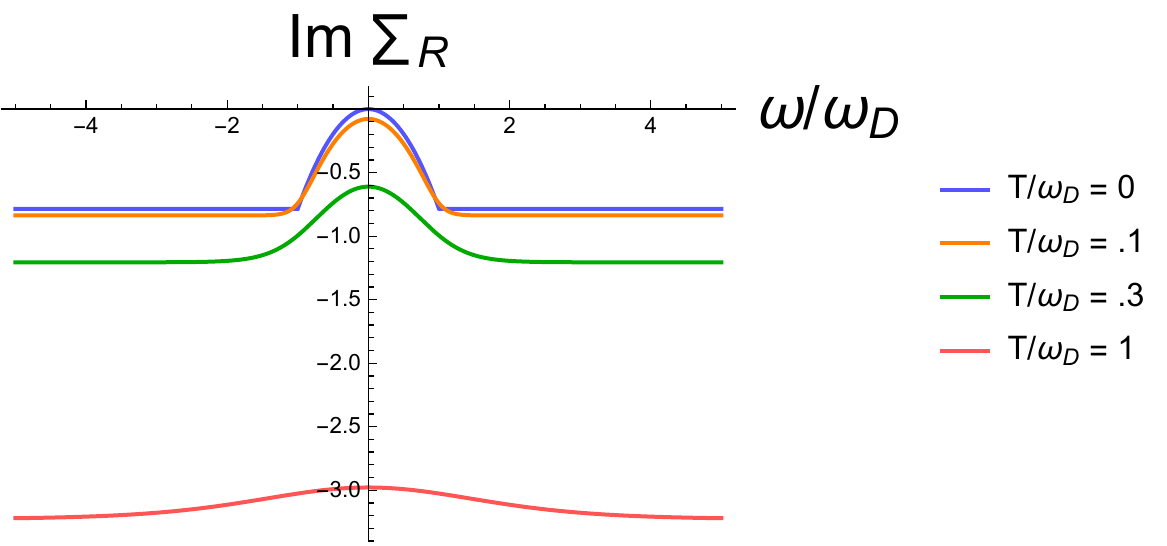}
	\caption{Plot of $\Sigma_R/\omega_D$ versus $\omega/\omega_D$ with $T$ held constant for the Debye phonon model in 2D, with $g^2\omega_Dk_0/(8\pi^2c\mu)=.25$.}
	\label{fig:d2sigmar}
\end{figure}

\begin{figure}[!htb]
	\centering
	\includegraphics[width=\columnwidth]{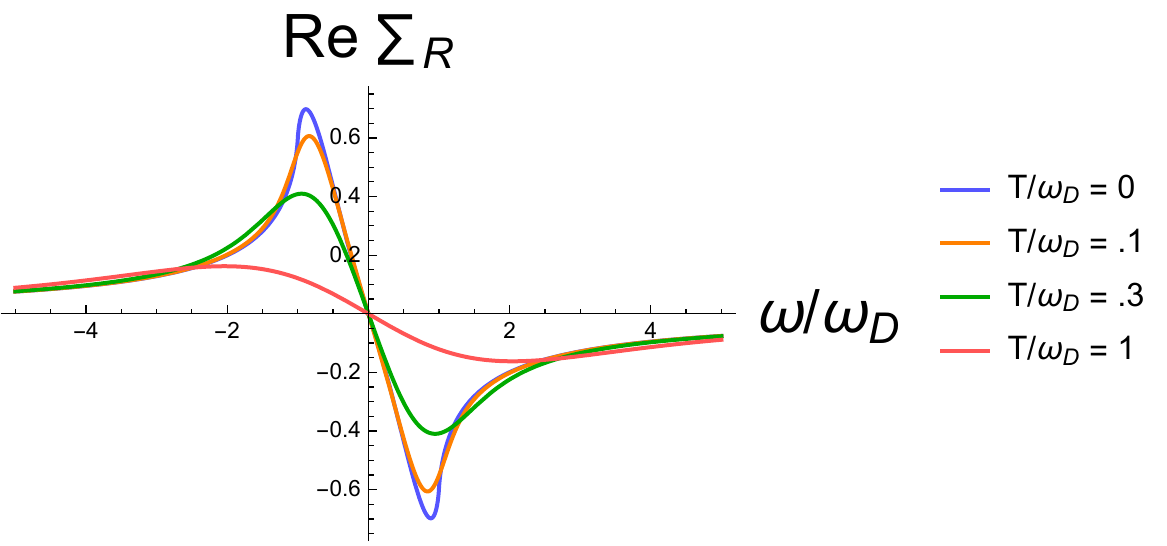}	
	\includegraphics[width=\columnwidth]{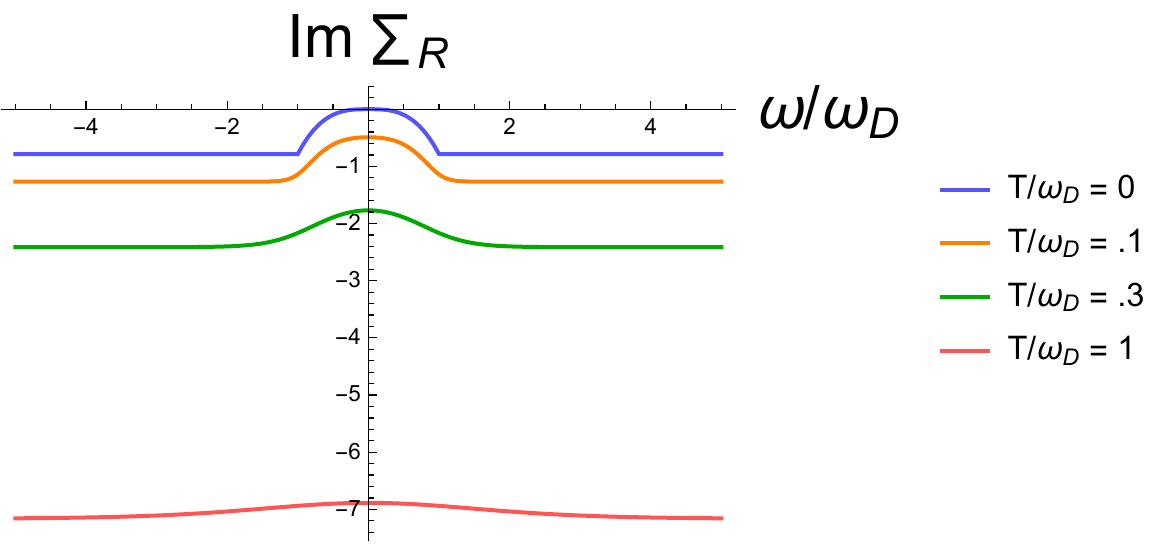}
	\caption{Plot of $\Sigma_R/\omega_D$ versus $\omega/\omega_D$ with $T$ held constant for the Debye phonon model in 3D, with $g^2\omega_D^2k_0/(24\pi^2c^2\mu)=.25$.}
\end{figure}

\begin{figure}[!htb]
	\centering
	\includegraphics[width=\columnwidth]{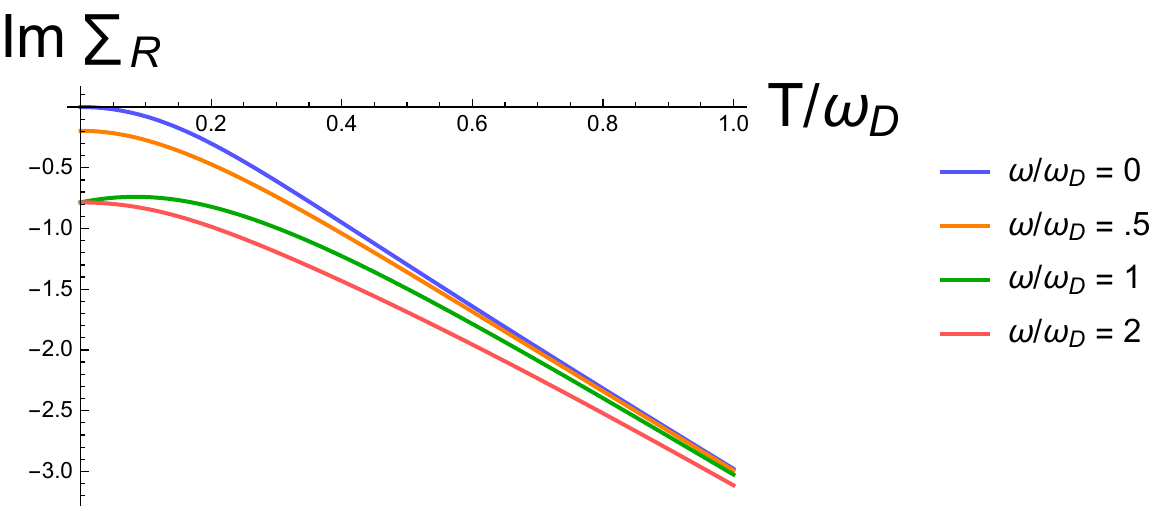}
	\caption{Plot of $\im\Sigma_R$ versus $T/\omega_0$ with $\omega$ held constant for the Debye phonon model in 2D, with $g^2\omega_Dk_0/(8\pi^2c\mu)=.25$.}
\end{figure}

\begin{figure}[!htb]
	\centering
	\includegraphics[width=\columnwidth]{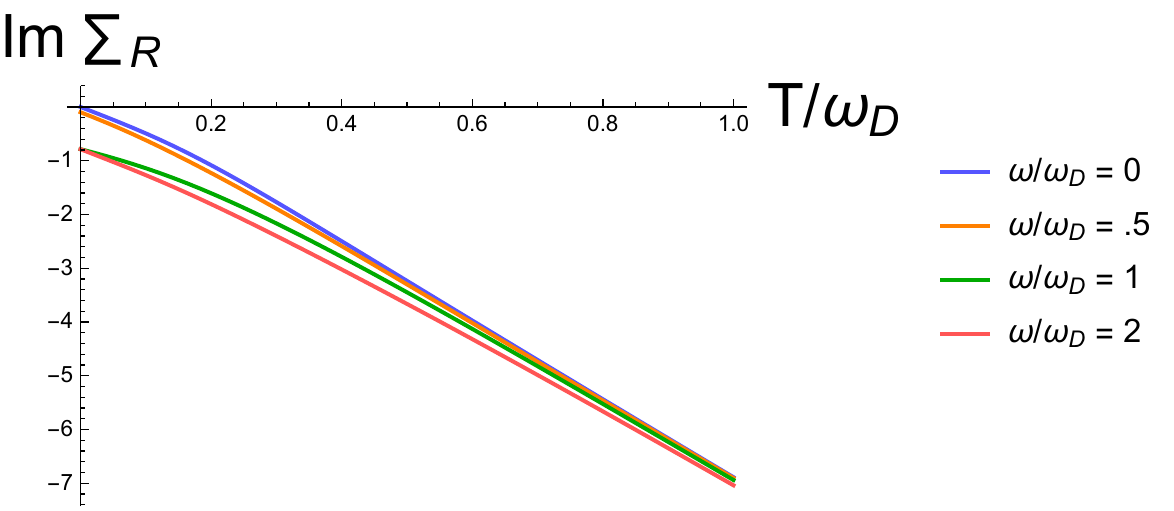}
	\caption{Plot of $\im\Sigma_R$ versus $T/\omega_0$ with $\omega$ held constant for the Debye phonon model in 3D, with $g^2\omega_D^2k_0/(24\pi^2c^2\mu)=.25$.}
	\label{fig:d3srt}
\end{figure}

\begin{figure*}[!htb]
	\centering
	\includegraphics[width=2\columnwidth]{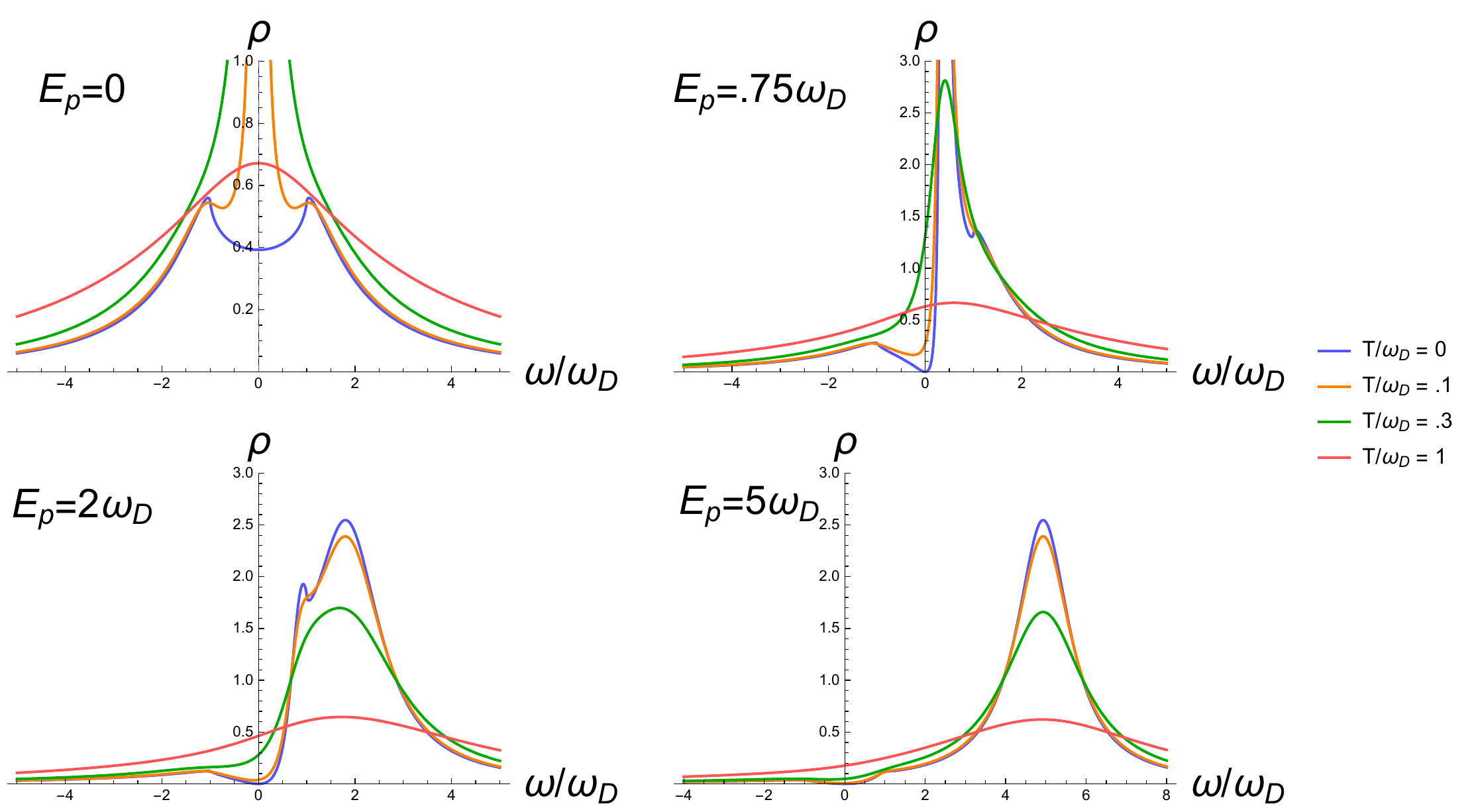}	
	\caption{Plots of $\rho(E_p,\omega)$ for different values of $T$ and $E_p$ for the Debye phonon model in 2D with $g^2\omega_Dk_0/(8\pi^2c\mu)=.25$.}
	\label{fig:d2rho}
\end{figure*}

Evaluating the sum and integral and analytically continuing to find the retarded self-energy yields the following:
\begin{align}
&\Sigma_R^{\text{2D}}(\omega)=\frac{g^2\omega_DTk_0}{2\pi c\mu}\Bigg(i\Bigg[\psi^{(-1)}\Big(i\frac{\omega_D}{2\pi T}\Big)+\psi^{(-1)}\Big(-i\frac{\omega_D}{2\pi T}\Big)\nonumber\\&+1-\psi^{(-1)}\Big(\frac{1}{2}+i\frac{\omega_D-\omega}{2\pi T}\Big)-\psi^{(-1)}\Big(\frac{1}{2}+i\frac{-\omega_D-\omega}{2\pi T}\Big)\Bigg]\nonumber\\
&+\frac{2\pi T}{\omega_D}\Bigg[\psi^{(-2)}\Big(\frac{1}{2}+i\frac{\omega_D-\omega}{2\pi T}\Big)-\psi^{(-2)}\Big(\frac{1}{2}+i\frac{-\omega_D-\omega}{2\pi T}\Big)\nonumber\\&\qquad-\psi^{(-2)}\Big(i\frac{\omega_D}{2\pi T}\Big)+\psi^{(-2)}\Big(-i\frac{\omega_D}{2\pi T}\Big)\Bigg]\Bigg)
\end{align}

\begin{figure*}[!htb]
	\centering
	\includegraphics[width=2\columnwidth]{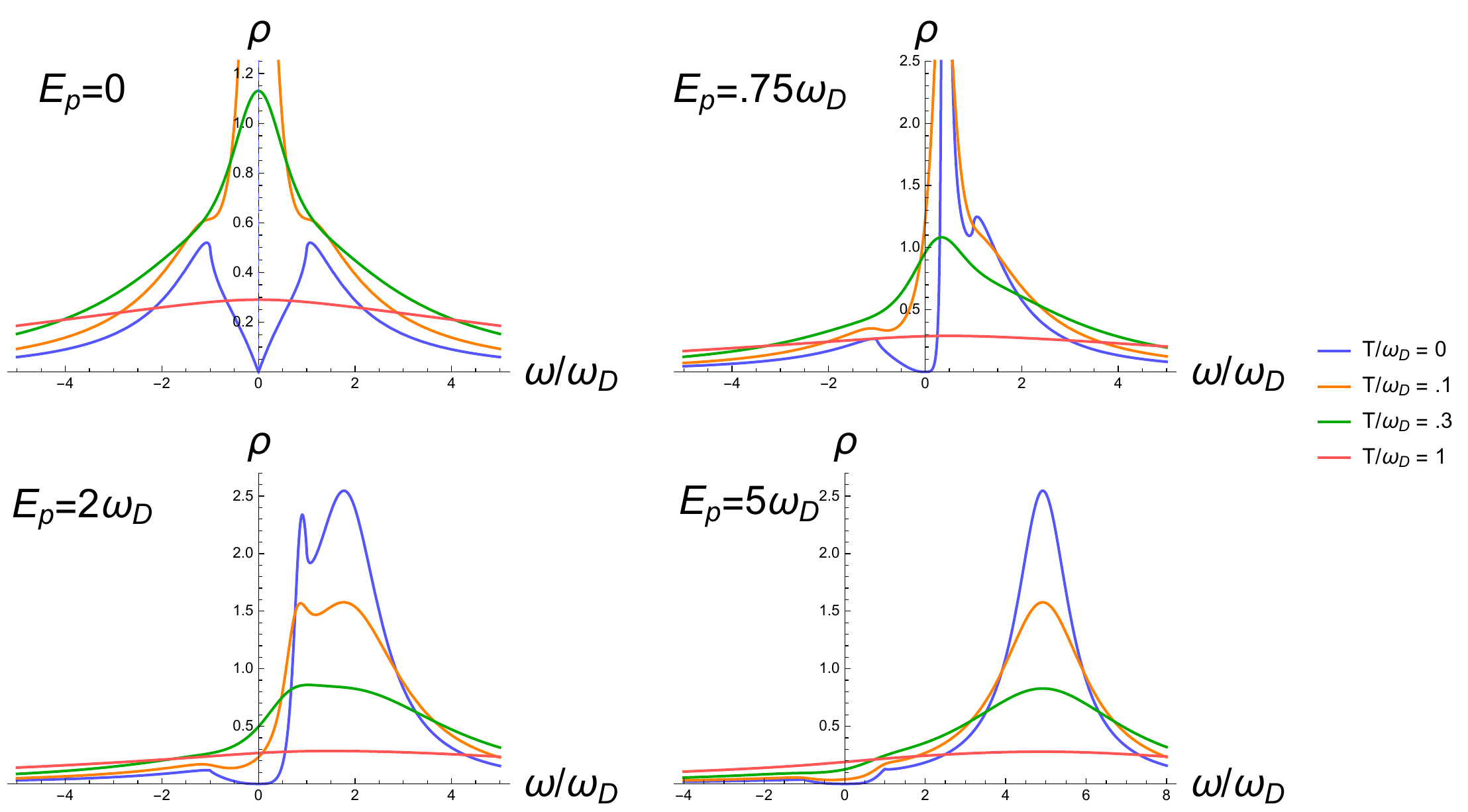}	
	\caption{Plots of $\rho(E_p,\omega)$ for different values of $T$ and $E_p$ for the Debye phonon model in 3D with $g^2\omega_D^2k_0/(24\pi^2c^2\mu)=.25$.}
	\label{fig:d3rho}
\end{figure*}

\begin{align}
&\Sigma_R^{\text{3D}}(\omega)=\frac{g^2T\omega_D^2k_0}{8\pi c^2\mu}\Bigg(i\frac{16\pi^2T^2}{\omega_D^2}\Bigg[-2\psi^{(-3)}\Big(\frac{1}{2}-i\frac{\omega}{2\pi T}\Big)\nonumber\\&-\psi^{(-3)}\Big(i\frac{\omega_D}{2\pi T}\Big)-\psi^{(-3)}\Big(-i\frac{\omega_D}{2\pi T}\Big)\nonumber\\&+\psi^{(-3)}\Big(\frac{1}{2}+i\frac{\omega_D-\omega}{2\pi T}\Big)+\psi^{(-3)}\Big(\frac{1}{2}+i\frac{-\omega_D-\omega}{2\pi T}\Big)\Bigg]\nonumber\\&+2i\Bigg[\psi^{(-1)}\Big(i\frac{\omega_D}{2\pi T}\Big)+\psi^{(-1)}\Big(-i\frac{\omega_D}{2\pi T}\Big)\nonumber\\&-\psi^{(-1)}\Big(\frac{1}{2}+i\frac{\omega_D-\omega}{2\pi T}\Big)-\psi^{(-1)}\Big(\frac{1}{2}+i\frac{-\omega_D-\omega}{2\pi T}\Big)-\frac{1}{2}\Bigg]\nonumber\\&+\frac{8\pi T}{\omega_D}\Bigg[\psi^{(-2)}\Big(\frac{1}{2}+i\frac{\omega_D-\omega}{2\pi T}\Big)-\psi^{(-2)}\Big(\frac{1}{2}+i\frac{-\omega_D-\omega}{2\pi T}\Big)\nonumber\\&-\psi^{(-2)}\Big(i\frac{\omega_D}{2\pi T}\Big)+\psi^{(-2)}\Big(-i\frac{\omega_D}{2\pi T}\Big)\Bigg]\Bigg)
\end{align}

We show the calculated self-energy in the Debye model for different parameters in figs. \ref{fig:d2sigmar}-\ref{fig:d3srt}.  The real part of the self-energy is linear in frequency for small frequencies as in the Einstein model, and the imaginary part again manifests a linear-in-$T$ behavior similar to the Einstein model provided $T>0.2 \omega_D$.  Thus, again the system manifests an effective marginal Fermi liquid behavior for $T>\omega_D/5$, which could be low if the relevant phonon modes are of low energies.  Again, a linear-in-$T$ resistivity would persist down to $T\sim \omega_D/5$.\cite{HwangPRB2019} -- note that although this is asymptotically a high-temperature behavior for the coupled electron-phonon system, experimentally the linearity could persist to low absolute temperatures if the relevant phonon modes are soft.

\begin{figure}[!htb]
	\centering
	\includegraphics[width=\columnwidth]{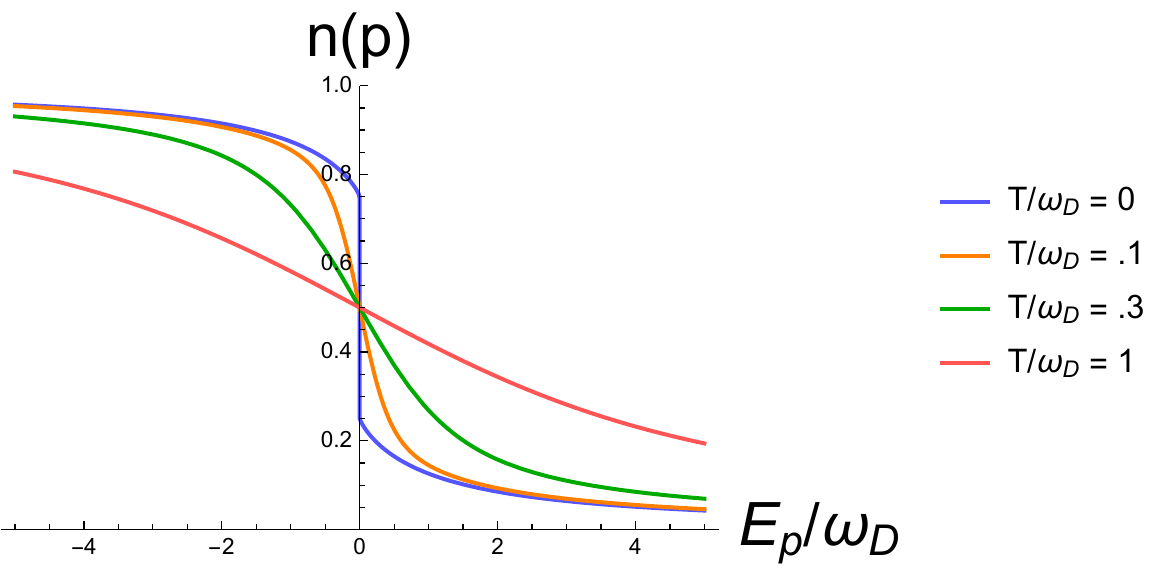}	
	\caption{Plot of $n(p)$ for different values of $T$ for the Debye phonon model in 2D, with $g^2\omega_Dk_0/(8\pi^2c\mu)=.25$.}
	\label{fig:d2n}
\end{figure}

\begin{figure}[!htb]
	\centering
	\includegraphics[width=\columnwidth]{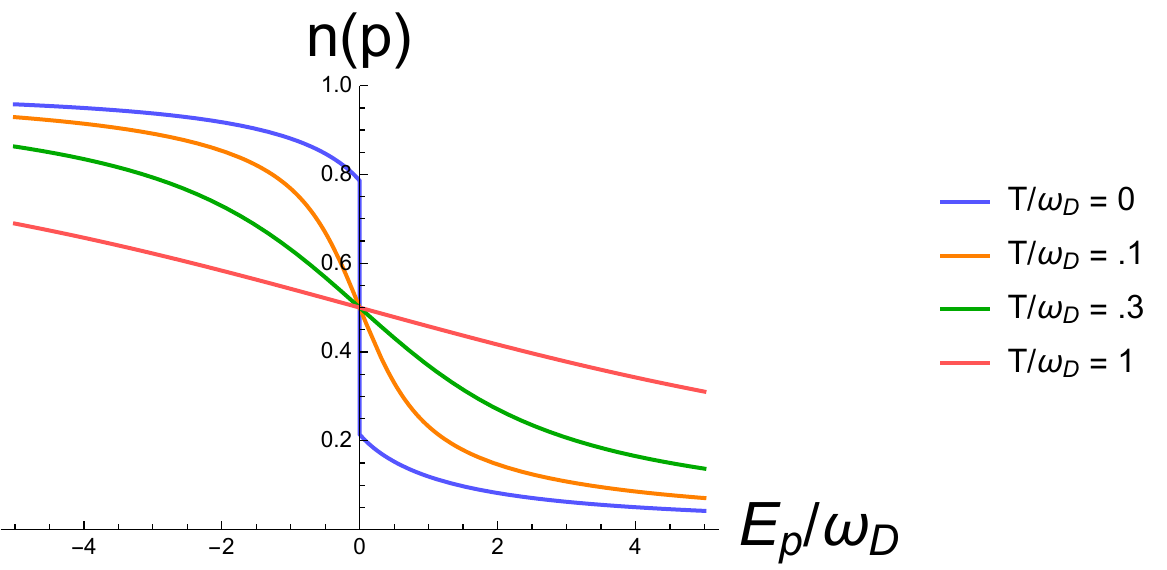}	
	\caption{Plot of $n(p)$ for different values of $T$ for the Debye phonon model in 3D, with $g^2\omega_D^2k_0/(24\pi^2c^2\mu)=.25$.}
	\label{fig:d3n}
\end{figure}

In figs. \ref{fig:d2rho} and \ref{fig:d3rho}, we show the calculated electronic spectral function, and in figs. \ref{fig:d2n} and \ref{fig:d3n}, the calculated momentum distribution function, for different sets of parameters in the Debye model using the calculated self-energies of figs. \ref{fig:d2sigmar}-\ref{fig:d3srt}.  Below we discuss these results.

\subsection{Expansions and Discussion}

Expanding for $T\ll\omega_D,\omega_D\pm\omega$, we obtain:
\begin{align}
&\Sigma_R^{\text{2D}}(\omega)=\frac{g^2}{8\pi^2 cv_F}\Bigg(-\pi i\omega_D^2+\Big(\omega_D^2-\omega^2-\frac{\pi^2}{3}T^2\Big)\Bigg[\nonumber\\&\pi i\Theta(\omega_D^2-\omega^2)-\frac{2\omega\omega_D}{\omega_D^2-\omega^2}+\log\bigg|\frac{\omega-\omega_D}{\omega+\omega_D}\bigg|\Bigg]\nonumber\\&-\frac{2\pi^3}{3}iT^2+O\big(T^3\big)\Bigg)
\end{align}

\begin{align}
&\Sigma_R^{\text{3D}}(\omega)=\frac{g^2}{24\pi^2 c^2v_F}\Bigg(-\pi i\omega_D^3\Theta(\omega^2-\omega_D^2)\nonumber\\&-\pi i\omega^3\sgn\omega\Theta(\omega_D^2-\omega^2)-\omega_D^2\omega+\omega^3\log\bigg|\frac{\omega^2}{\omega^2-\omega_D^2}\bigg|\nonumber\\&+\omega_D^3\log\bigg|\frac{\omega-\omega_D}{\omega+\omega_D}\bigg|-6\pi i\omega_D^2T+O\big(T^2\big)\Bigg)
\end{align}

In particular, we note that for low $T$, the corrections to $\im\Sigma_R$ are linear in $T$ for the 3D case, in comparison to the 2D case and the Einstein model, which are quadratic in $T$. This causes the spectral function to broaden much more quickly as $T$ increases in three dimensions than in two. We plot the momentum distribution function $n(p)$, and find that similarly to the Einstein phonon model, there is a discontinuity at $p=k_F$ at zero temperature, and thus a Fermi surface exists (here $k_F$ is the Fermi momentum defined as $k_F=(2mE_F)^{1/2}$ with $E_F$ being the chemical potential of the noninteracting system at $T=0$). We note, however, that for low-temperatures, it does not approach a complete step function from 1 to 0, as not all the weight of the spectral function is contained within the delta function. Thus, electron-phonon interactions can lead to apparent NFL behaviors. The difference between the Debye model in 2D and 3D also becomes evident, as the linear scaling in $T$ of $\im\sigma_R$ in 3D causes the jump around $E_p=0$ to become much less steep at low temperatures than that of the corresponding 2D case, which has quadratic scaling in $T$.

Performing the high temperature expansion, where $T\gg\omega_D,\omega$, we obtain:

\begin{align}
\Sigma_R^{\text{2D}}(\omega)=\frac{g^2\omega_D^2k_0}{2\pi c\mu}\Bigg[&\,i\,\bigg(\frac{-T}{\omega_D}+\frac{\omega_D}{18T}\bigg)\nonumber\\&+\frac{\psi^{(2)}(\frac{1}{2})\omega\omega_D}{12\pi^3T^2}+O\bigg(\frac{1}{T^3}\bigg)\Bigg]
\end{align}

\begin{align}
\Sigma_R^{\text{3D}}(\omega)=\frac{g^2\omega_D^3k_0}{8\pi c^2\mu}\Bigg[&\,i\,\bigg(\frac{-3T}{\omega_D}+\frac{\omega_D}{12T}\bigg)\nonumber\\&+\frac{\psi^{(2)}(\frac{1}{2})\omega\omega_D}{8\pi^3T^2}+O\bigg(\frac{1}{T^3}\bigg)\Bigg]
\end{align}

Thus for the Debye model in both two and three dimensions, the high temperature behavior is similar to the Einstein model, with $\re\Sigma_R$ approaching 0 as $T^{-2}$, and with $\im\Sigma_R$ becoming linear in $T$, with the only difference in each case being the exact coefficients. We emphasize that this apparent high-temperature behavior, however, already manifests itself at $T\sim \omega_D/5$, indicating that the marginal Fermi liquid properties could be apparent in a coupled electron-phonon system down to $T\sim \omega_D/5$. We treat these cases together by defining $\kappa$ such that $\im\Sigma_R\rightarrow-\kappa g^2T$. Then $\kappa$ is given by:
\begin{equation}
\kappa=
\begin{cases}
\pi N_d&\text{for Einstien phonons in all}\\
&\qquad\qquad\qquad\;\;\;\text{dimensions}\\
\omega_Dk_0/(2\pi c\mu)&\text{for Debye phonons in 2D}\\
3\omega_D^2k_0/(8\pi c^2\mu)&\text{for Debye phonons in 3D}
\end{cases}
\end{equation}

Then for large $T$, the spectral function approaches a Lorentzian with half-width $\kappa g^2T$:
\begin{equation}
\rho(\omega,p)=\frac{2\kappa g^2T}{(\omega-E_p)^2+(\kappa g^2T)^2}
\end{equation}

We emphasize that this Lorentzian reflects an incoherent quasiparticle at finite temperatures, which could be very broad for strong electron-phonon coupling. From this, it is simple to analytically compute the momentum distribution function:
\begin{align}
n(p)=&\frac{i}{2\pi}\Bigg[\psi^{(0)}\bigg(\frac{1}{2}+\frac{\kappa g^2}{2\pi}+i\frac{E_p}{2\pi T}\bigg)\nonumber\\&-\psi^{(0)}\bigg(\frac{1}{2}-\frac{\kappa g^2}{2\pi}+i\frac{E_p}{2\pi T}\bigg)\Bigg]+\Big(1+e^{\kappa g^2 i+E_p/T}\Big)^{-1}
\label{eqn:nlarget}
\end{align}

\begin{figure}[!h]
	\centering
	\includegraphics[width=\columnwidth]{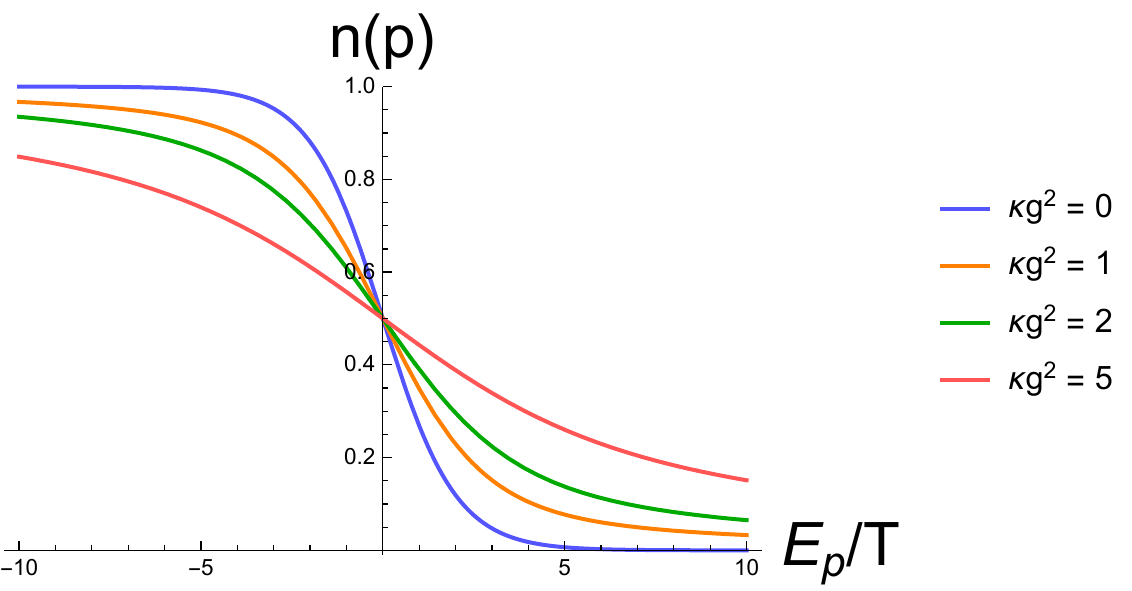}	
	\caption{High temperature limit of $n(p)$ for different values of $\kappa g^2$.}
	\label{fig:nlarget}
\end{figure}

Specifically, we note that the high temperature momentum distribution function depends only on the coupling strength $\kappa g^2$ and the ratio $E_p/T$. It is easy to see that as the coupling strength goes to zero, the standard Fermi-Dirac distribution is obtained. In fig. \ref{fig:nlarget} we plot the calculated high-$T$ form of the momentum distribution function for the Debye model based on Eq. (\ref{eqn:nlarget}).

The high-temperature momentum-distribution function at first appears to be qualitatively similar to the Fermi-Dirac distribution; however, there is an important distinction between the two. As $E_p/T\rightarrow\infty$, $n(p)\rightarrow\kappa g^2\pi^{-1}(E_p/T)^{-1}$, rather than dropping off exponentially in $(E_p/T)$ as does the Fermi-Dirac distribution. Thus, although the interesting features of the spectral function get smoothed out at high temperatures, this departure from standard Fermi-Dirac statistics still provides an example of NFL behavior. The coupled electron-phonon system thus provides a simple analytically tractable effective high-temperature NFL model.

\section{Electron-Impurity Interactions}

\subsection{Self-Energy Calculation}

We now consider an electron gas with random quenched impurities in $d$ dimensions. We again assume that the dispersion takes the form $E_p=p^2/2m$, and we include an electron-impurity potential given by Thomas-Fermi screening of a point charge as follows:
\begin{equation}
u(q)=\frac{u_0}{1+(q/q_{TF})^{d-1}}
\end{equation}

where $q_{TF}$ is the Thomas-Fermi wavevector, and $u_0$ dictates the strength of the interaction. Following Ref. \onlinecite{DasSarmaPRB1981}, to lowest order, the ensemble-averaged self-energy is given by the Born approximation as:
\begin{equation}
\Sigma(p,i\omega_n)=N_i\int\frac{d^dk}{(2\pi)^d}\;u(\vec{k}-\vec{p})^2\;G_0^*(k,i\omega_n)
\label{eqn:sefreeg}
\end{equation}

where $N_i$ is the impurity concentration. Here we use the free Green's function with renormalized parameters:
\begin{equation}
G_0^*(p,i\omega_n)=\frac{Z}{i\omega_n-p^2/2m^*+\mu^*}
\end{equation}

\begin{figure}[!htb]
	\centering
	\includegraphics[width=\columnwidth]{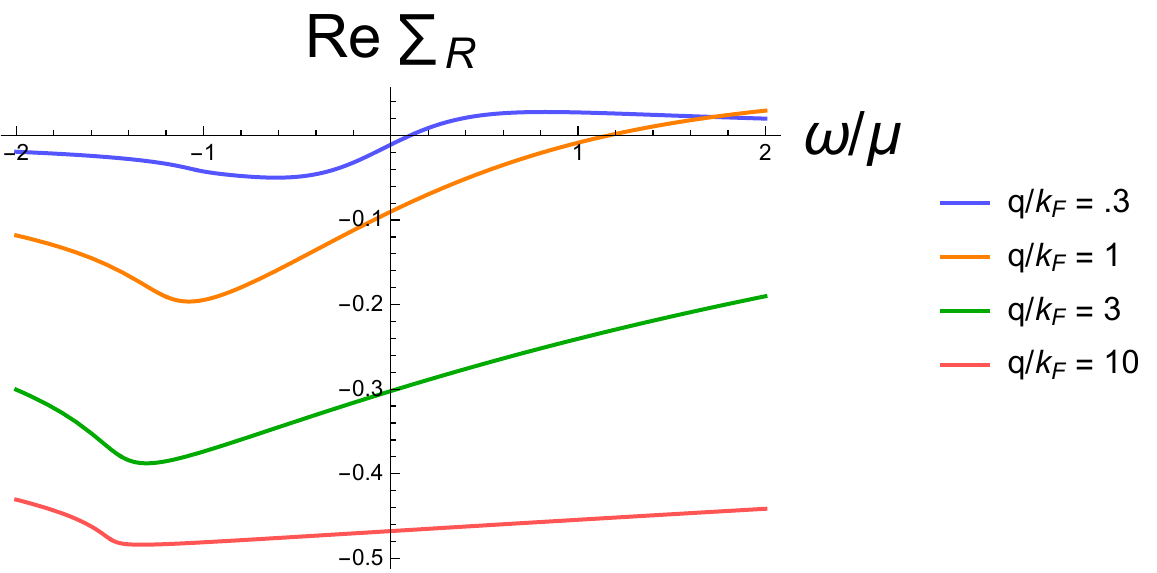}	
	\includegraphics[width=\columnwidth]{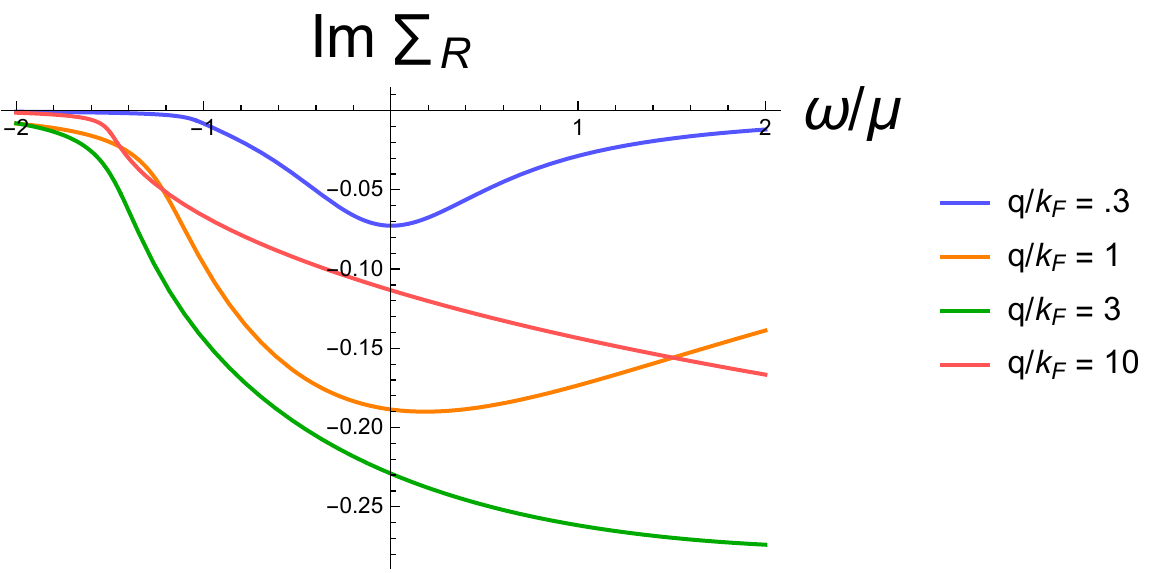}
	\caption{Plot of $\Sigma_R(\omega)$ for different values of $q_{TF}$ with $p=k_F$ and $\alpha=.5$ for the electron-impurity model in 3D.}
	\label{fig:i3sr}
\end{figure}

In three dimensions, standard techniques can be used to evaluate the integral. However, in two dimensions, the non-analyticity of the Thomas-Fermi screening potential (due to its dependence on $q$ rather than $q^2$) makes an analytical evaluation of the integral nearly impossible. Thus, in this paper we will give analytical expressions for the 3D case, and provide numerical results for the 2D case.

Then in three dimensions, we get after carrying out the integrals:

\begin{align}
&\Sigma_{\text{3D}}(p,\omega+i0)=\frac{-Zm^*N_iu_0^2q_{TF}}{4\pi}\times\nonumber\\&\frac{q_{TF}^4}{(q_{TF}^2+p^2+2m^*(\mu^*+\omega))^2-8p^2m^*(\mu^*+\omega)}\times\nonumber\\&\bigg(\frac{q_{TF}^2+p^2-2m^*(\mu^*+\omega)}{q_{TF}^2}+\frac{2i\sqrt{2m^*(\mu^*+\omega)}}{q_{TF}}\bigg)
\end{align}

\begin{figure}[!htb]
	\centering
	\includegraphics[width=\columnwidth]{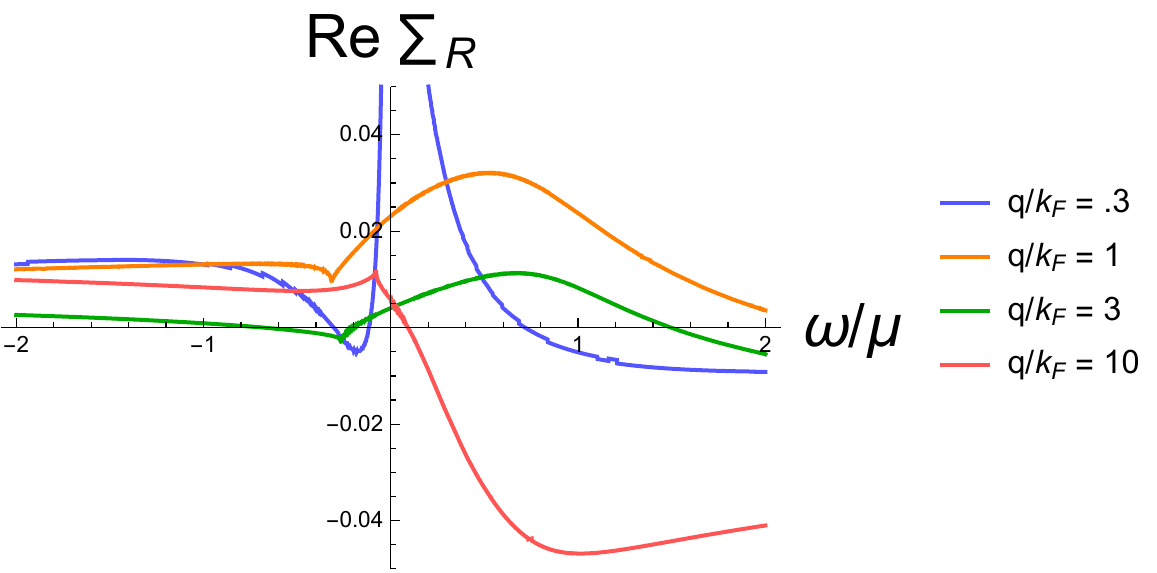}	
	\includegraphics[width=\columnwidth]{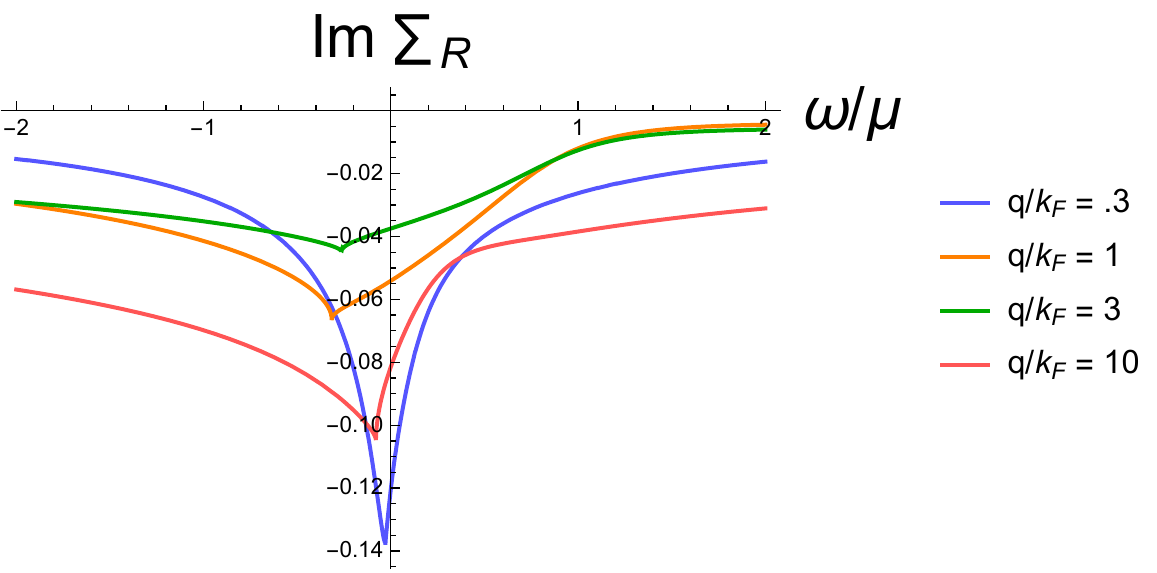}
	\caption{Plot of $\Sigma_R(\omega)$ for different values of $q_{TF}$ with $p=k_F$ and $\alpha=2$ for the electron-impurity model in 2D.}
	\label{fig:i2sr}
\end{figure}

$Z$, $m^*$, and $\mu^*$ are then found by setting $G(p,\omega)=G_0^*(p,\omega)$ after Taylor expanding $\Sigma$ to first order about the Fermi energy and momentum:
\begin{equation}
\Sigma(p,\omega)\approx \Sigma(k_F,0)+\omega\frac{d\Sigma}{d\omega}\big(k_F,0\big)+(p^2-k_F^2)\frac{d\Sigma}{d(p^2)}\big(k_F,0\big)
\end{equation}
which gives the following for $Z$, $m^*$, and $\mu^*$:

\begin{flalign}
&Z^{-1}=1-\frac{d\Sigma}{d\omega}\big(k_F,0\big)&&\nonumber\\
&=1+\frac{mN_iu_0^2q_{TF}}{4\pi}\;\frac{2m(-3q_{TF}^2-8m\mu+iq_{TF}^3/k_F)}{(q_{TF}^2+8m\mu)^2}&&
\end{flalign}
\begin{flalign}
\frac{1}{2m^*}&=Z\frac{1}{2m}+\frac{d\Sigma}{dp^2}\big(k_F,0\big)&&\nonumber\\
&=Z\frac{1}{2m}+\frac{mN_iu_0^2q_{TF}}{4\pi}\;\frac{q_{TF}^2-8m\mu}{(q_{TF}^2+8m\mu)^2}&&
\end{flalign}
\begin{flalign}
\mu^*&=Z\mu-\Sigma(k_F,0)+k_F^2\frac{d\Sigma}{dp^2}\big(k_F,0\big)&&\nonumber\\
&=Z\mu+\frac{mN_iu_0^2q_{TF}}{4\pi}\times&&\nonumber\\&\frac{q_{TF}^4+10q_{TF}^2m\mu-16m^2\mu^2+2iq_{TF}k_F(q_{TF}^2+8m\mu)}{(q_{TF}^2+8m\mu)^2}&&
\end{flalign}

\begin{figure}[!htb]
	\centering
	\includegraphics[width=\columnwidth]{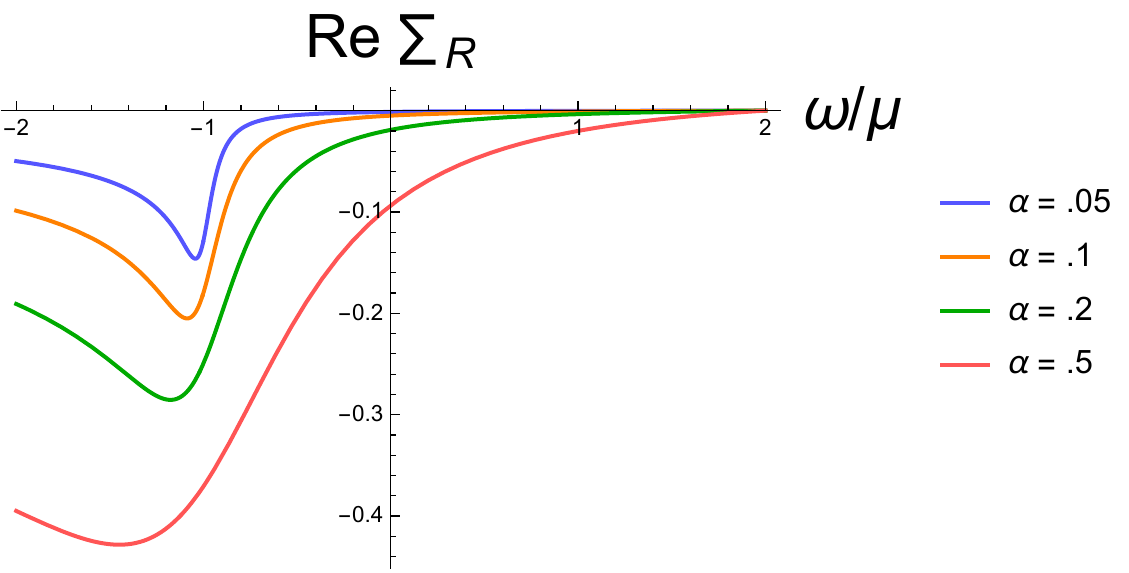}	
	\includegraphics[width=\columnwidth]{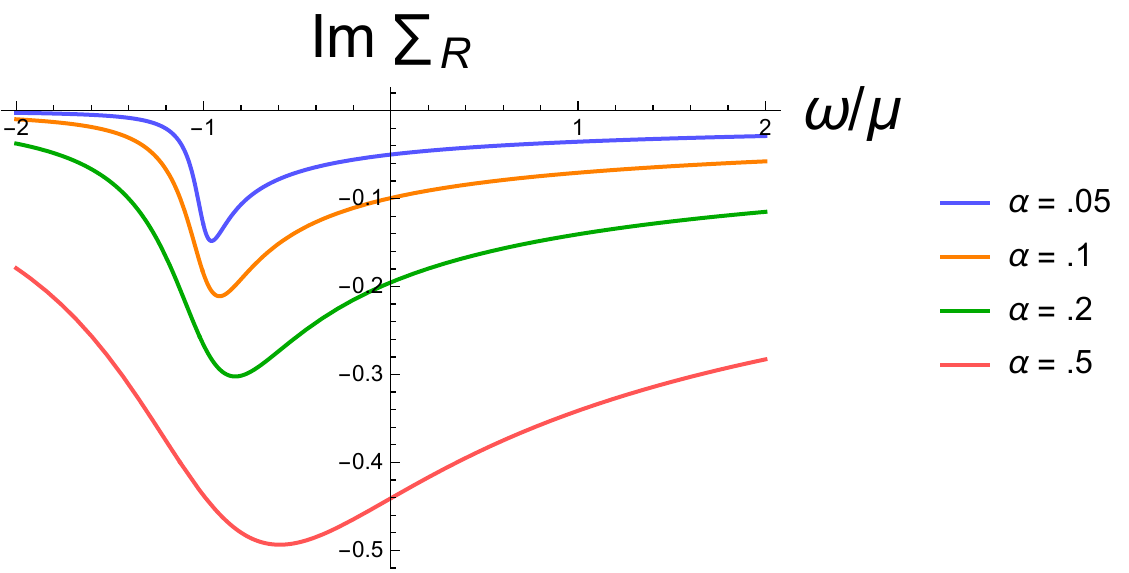}
	\caption{Plot of $\Sigma_R(\omega)$ for the electron-impurity model in 1D.}
	\label{fig:i1sr}
\end{figure}

For completeness, we also consider a 1D electron gas (non-interacting, so as to avoid all Luttinger liquid complications) with constant electron-impurity potential $u(q)=u_0/2$. Then the retarded self-energy is simply given by:

\begin{equation}
\Sigma_{\text{1D}}(\omega+i0)=\frac{-Nu_0^2m^*Zi}{4\sqrt{2m^*(\omega+\mu^*)}}
\end{equation}

\begin{figure}[!htb]
	\centering
	\includegraphics[width=\columnwidth]{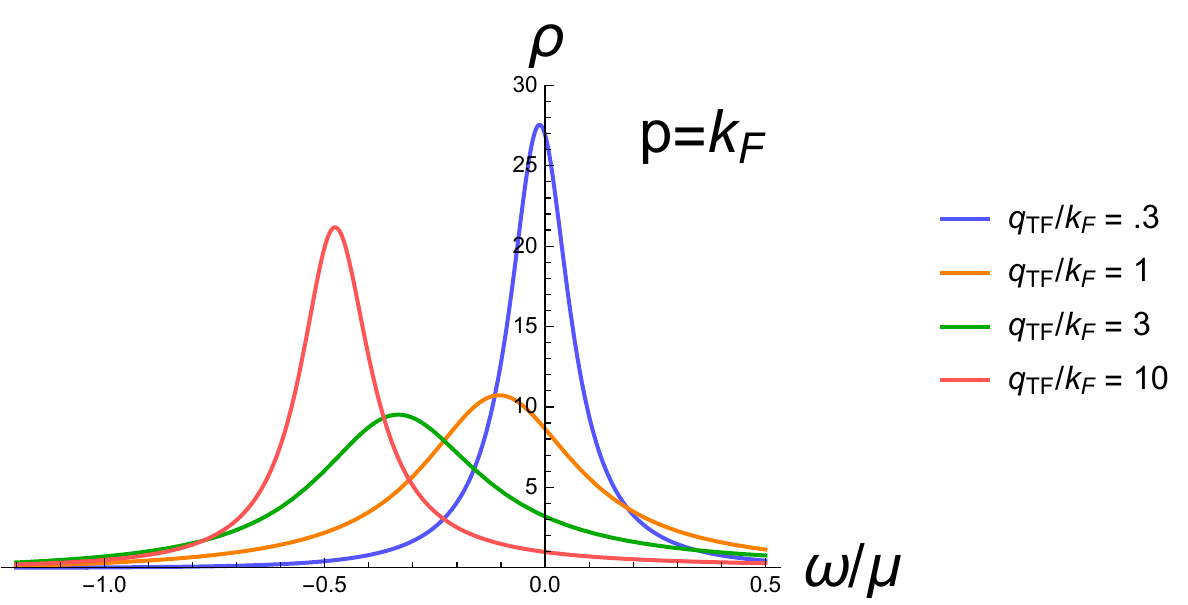}	
	\caption{Plot of $\rho(p,\omega)$ for different values of $q_{TF}$ with $p=k_F$ and $\alpha=.5$ for the electron-impurity model in 3D.}
	\label{fig:i3rhoq}
\end{figure}

\begin{figure}[!htb]
	\centering
	\includegraphics[width=\columnwidth]{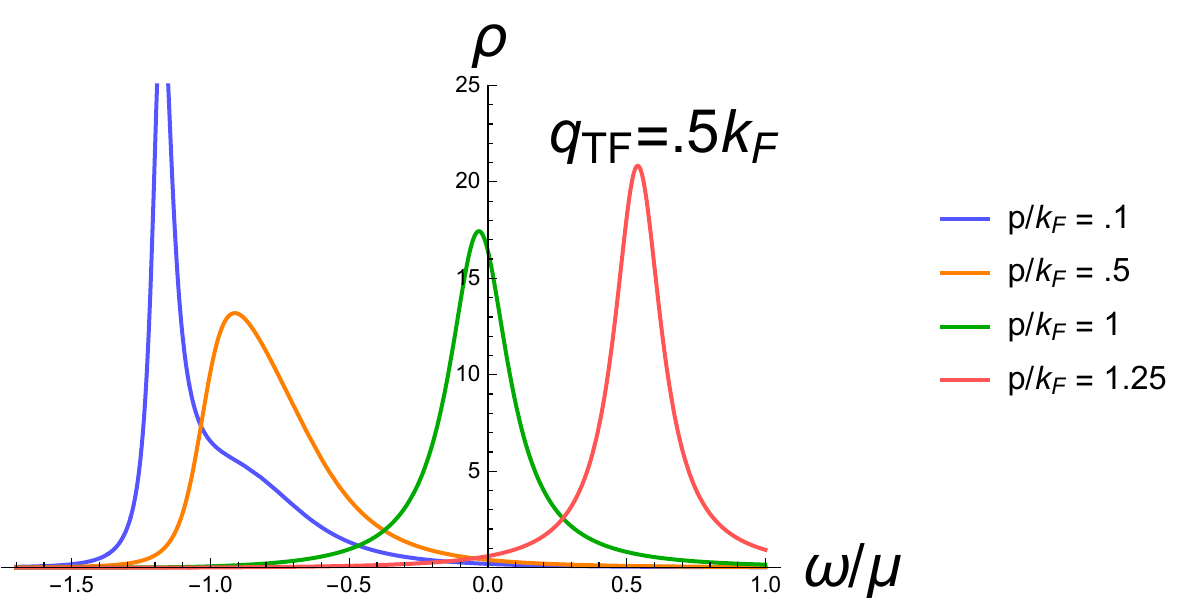}	
	\caption{Plot of $\rho(p,\omega)$ for different values of $p$ with $q_{TF}=.5k_F$ and $\alpha=.5$ for the electron-impurity model in 3D.}
	\label{fig:i3rhop}
\end{figure}

\begin{figure}[!htb]
	\centering
	\includegraphics[width=\columnwidth]{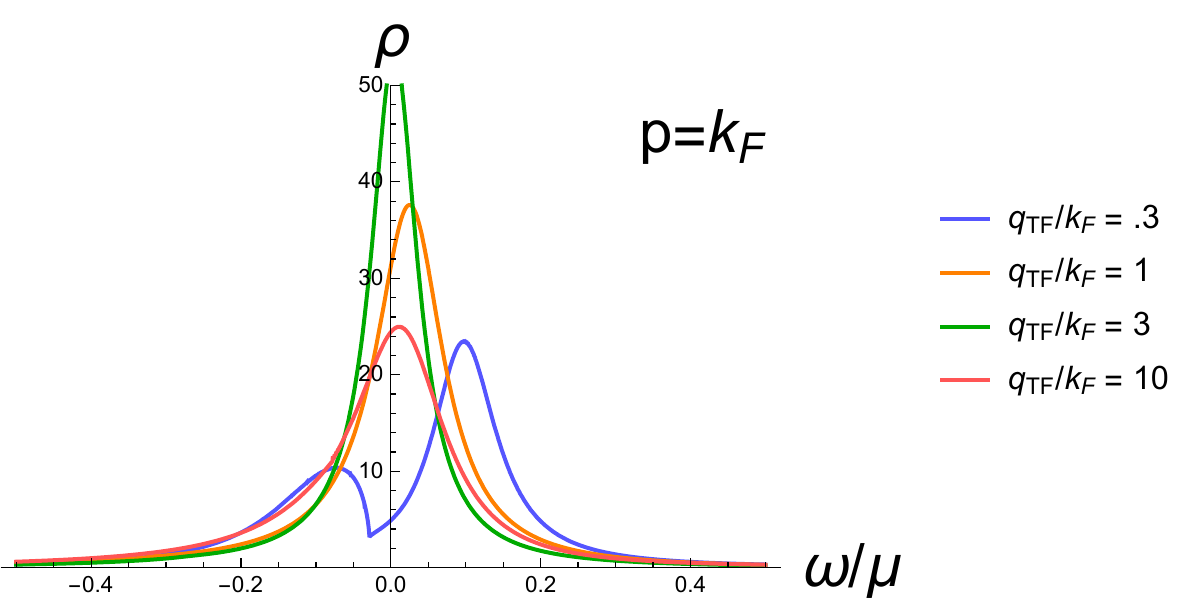}	
	\caption{Plot of $\rho(p,\omega)$ for different values of $q_{TF}$ with $p=k_F$ and $\alpha=2$ for the electron-impurity model in 2D.}
	\label{fig:i2rhoq}
\end{figure}

\begin{figure}[!htb]
	\centering
	\includegraphics[width=\columnwidth]{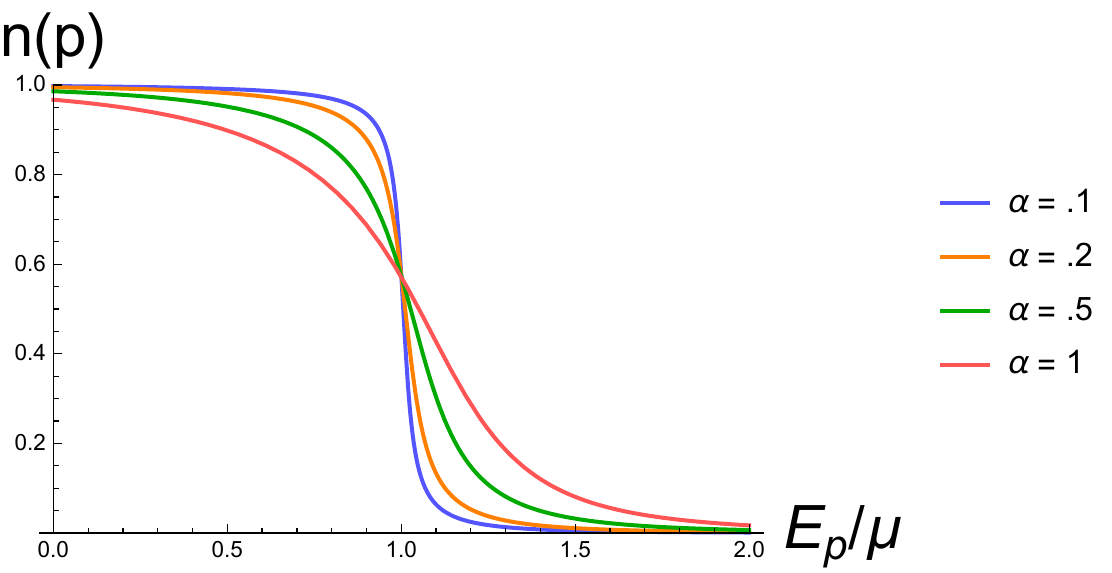}	
	\caption{Plot of $n(p)$ versus $p^2/k_F^2$ at zero temperature for different values of $\alpha$ and with $q_{TF}=.5k_F$ for the electron-impurity model in 3D.}
	\label{fig:i3n}
\end{figure}

\begin{figure}[!htb]
	\centering
	\includegraphics[width=\columnwidth]{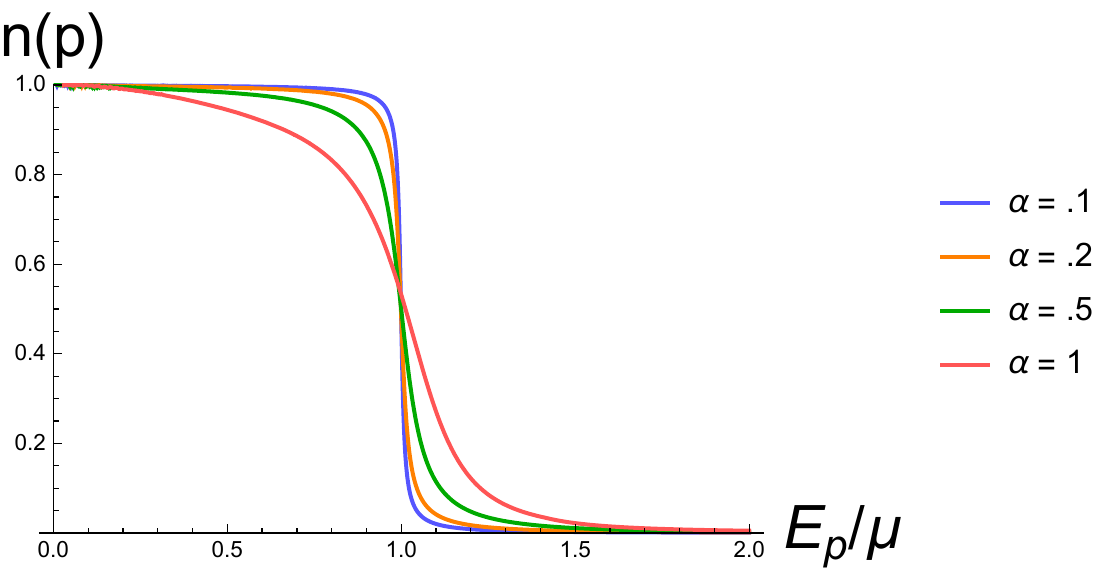}	
	\caption{Plot of $n(p)$ versus $p^2/k_F^2$ at zero temperature for different values of $\alpha$ and with $q_{TF}=.5k_F$ for the electron-impurity model in 2D.}
	\label{fig:i2n}
\end{figure}

\begin{figure}[!htb]
	\centering
	\includegraphics[width=\columnwidth]{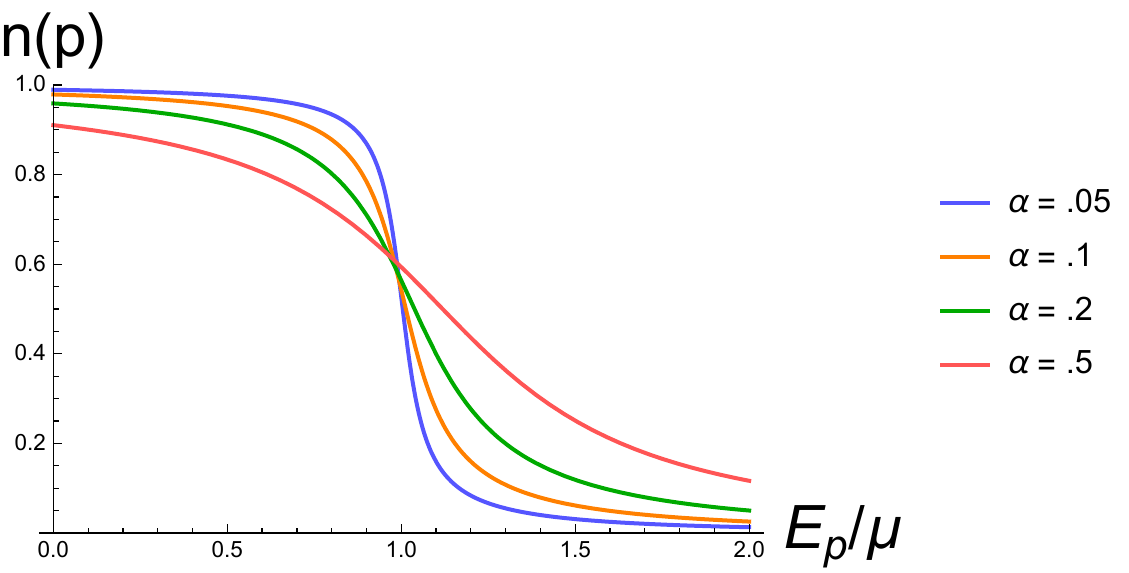}	
	\caption{Plot of $n(p)$ versus $p^2/k_F^2$ at zero temperature for different values of $\alpha$ for the electron-impurity model in 1D.}
	\label{fig:i1n}
\end{figure}

The calculated self-energies for 3D, 2D, and 1D systems are shown respectively in figs. \ref{fig:i3sr}, \ref{fig:i2sr}, and \ref{fig:i1sr}.  Using the self-energies, we then calculate the corresponding spectral functions, as shown in figs. \ref{fig:i3rhoq}/\ref{fig:i3rhop} (for 3D) and \ref{fig:i2rhoq} (for 2D), as well as the momentum distribution functions in 3D (fig. \ref{fig:i3n}), 2D (fig. \ref{fig:i2n}), and 1D (fig. \ref{fig:i1n}). 

\subsection{Discussion}

We note that the width of the spectral function scales roughly linearly with $q_{TF}$, and therefore define the dimensionless parameter $\alpha$ as: 

\begin{align}
\alpha_{\text{1D}}&=\frac{N_iu_0^2m}{4\mu k_F}\nonumber\\
\alpha_{\text{2D}}&=\frac{N_iu_0^2q_{TF}k_F}{\mu^2}\nonumber\\
\alpha_{\text{3D}}&=\frac{N_iu_0^2mq_{TF}}{4\pi\mu}
\end{align}


Because electron-impurity interactions give the self-energy a nonzero imaginary part for all momenta and energies, the zero-temperature spectral function has a finite width, and thus the momentum distribution function is continuous through the Fermi energy even at $T=0$. The lack of a discontinuity of $n(p)$ at the Fermi energy means that for a system with impurities, there is no well-defined Fermi surface by definition, i.e., impurities wash out the sharp discontinuity at $k=k_F$ in the momentum distribution function.  The system is thus trivially an NFL with no well-defined quasiparticles in momentum space. We can approximate the zero-temperature momentum distribution function in the presence of impurities with a Fermi-Dirac distribution at an effective temperature $T'$, by numerically minimizing the following integral:
\begin{equation}
\int_{0}^\infty \Big(n(E_p)-f(E_p-\mu',T')\Big)^2dE_p\\
\label{eqn:intminimpstd}
\end{equation}

\begin{figure}[!htb]
	\centering
	\includegraphics[width=\columnwidth]{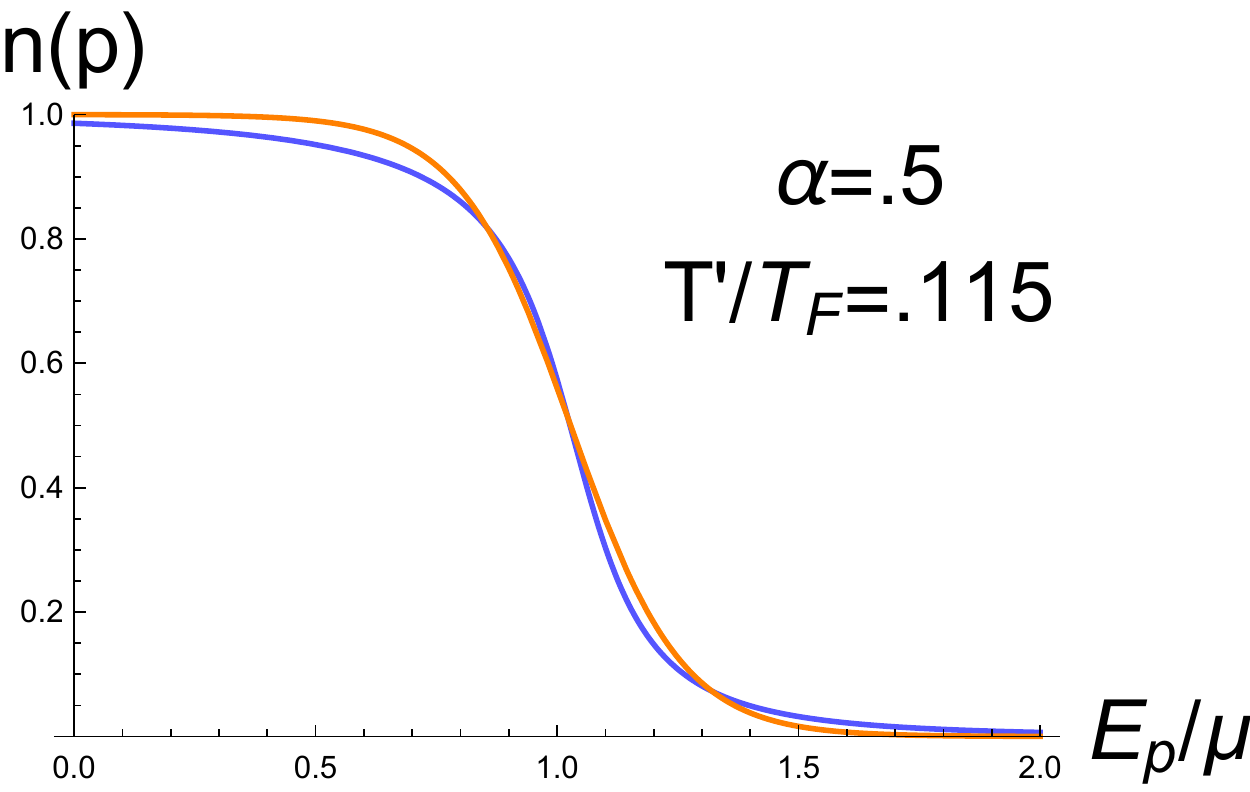}
	\caption{{\bf Blue:} The zero-temperature $n(p)$ for $q_{TF}=.5k_F$ and $\alpha=.5$ for the electron-impurity model in 3D. {\bf Orange:} The Fermi-Dirac distribution for an effective temperature $T'$ found by minimizing eq. \ref{eqn:intminimpstd}.}
\end{figure}

\begin{figure}[!htb]
	\centering
	\includegraphics[width=\columnwidth]{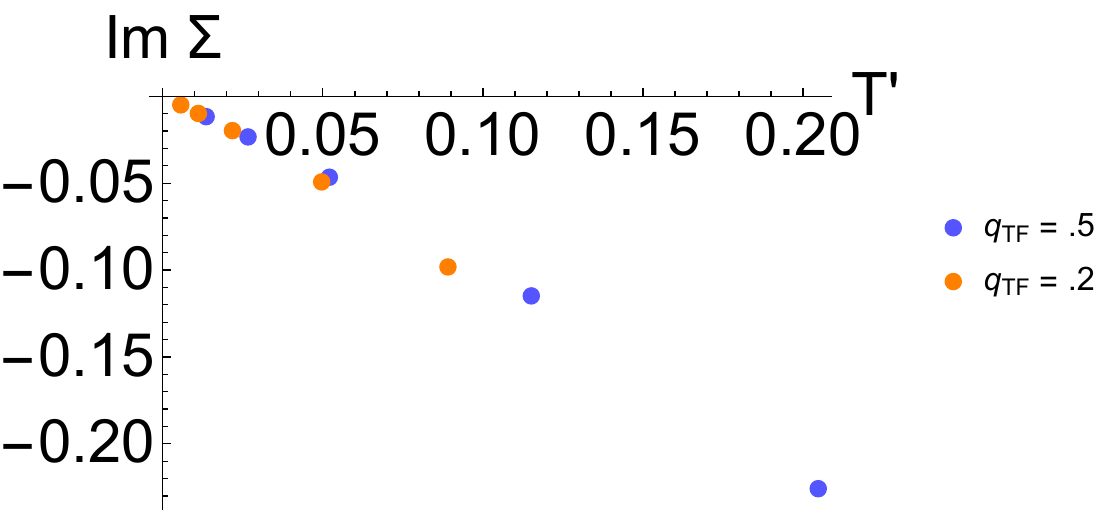}
	\caption{Plot of the effective temperature $T'$ versus $\im \Sigma_R(k_F,0)$ for values of $\alpha$ ranging from .05 to 1 for the electron-impurity model in 3D.}
\end{figure}

where $\mu'=\frac{\re m^*}{m}\re\mu^*$. We note that, like in the electron-phonon model discussed previously, $n(p)$ tends to approach 0 more slowly than a true Fermi-Dirac distribution as $E_p$ increases; however, the Fermi-Dirac distribution for an effective temperature $T'$ still appears to be a good approximation for $n(p)$ for $\alpha$ not too high. We note that the effective temperature $T'$ scales linearly with $\im \Sigma(k_F,0)$ as suggested by Dingle.\cite{DinglePRS1952} We find that the slope is given by $T'\approx.9|\im \Sigma|$, and is roughly independent of $q_{TF}$. Thus, the $T=0$ momentum distribution function in the presence of impurities simulates to some extent the finite temperature Fermi-Dirac distribution function at an effective temperature $T'$ without impurities.  Obviously, the momentum distribution function in the presence of impurities is continuous even at $T=0$, a hallmark of an NFL.  Here, the NFL behavior arises rather trivially by virtue of the fact that momentum is no longer conserved in the presence of impurity scattering leading to all momentum eigenstates being non-stationary and hence decaying even at $T=0$ with the decay given by the imaginary part of the self-energy.

\section{Conclusion}

We have calculated, to one loop order, the renormalized self-energy, spectral function, and momentum distribution function at finite temperature for the Einstein phonon model, the Debye phonon model, and the electron-impurity model. We have shown that simple mechanisms such as phonons or impurities cause systems to depart from the standard Fermi liquid paradigm, i.e., no discontinuity in the momentum distribution function (hence no quasiparticles) and incoherent spectral functions with no sharp features. In all phonon models, the spectral function has sharp, non-Lorentzian features at low temperatures, becoming bimodal in some cases. At high temperatures, the spectral function becomes Lorentzian, but with a width that scales linearly with $T$, which in turn causes the momentum distribution function to become much wider than the standard Fermi-Dirac distribution. In particular, $n(p)$ asymptoticly scales inversely with $E_p/T$ rather than dropping off exponentially as in a generic FL. Furthermore, we have demonstrated that electron-impurity interactions can also cause spectral widening at zero temperature, leading to a continuous momentum distribution function. This lack of discontinuity of $n(p)$ at zero temperature implies the lack of a well-defined Fermi surface.

Other than providing a simple demonstration of the apparent  NFL-like behavior in familiar contexts, where explicit calculations can be carried out using well-defined techniques, our work serves the purpose of a cautionary note to the invoking of the NFL paradigm whenever a measured spectral function manifests complicated behavior as a function of energy or a measured electronic resistivity manifests a linear-in-$T$ behavior down to some low temperature scale.  Our work shows that details matter, and unless the underlying Hamiltonian and the associated energy scales leading to an incoherent broad spectral function (or a linear-in-$T$ resistivity) are precisely known, it may be misleading to automatically associate such ``anomalous'' properties as manifesting an obvious non-Fermi-liquid ground state since it is possible that the apparent NFL behavior is arising trivially from the electrons coupling to very low energy phonons (or some other bosons) and/or quenched impurities in the system, providing an effective finite-temperature NFL behavior which will disappear at $T=0$ in the clean system in the absence of impurities.  Since clean systems at $T=0$ are often inaccessible experimentally, it is important to decisively rule out the manifestation of trivial NFL in every experimental context before a nontrivial breakdown of the quasiparticle picture can be asserted.

\acknowledgments
This work is supported by the Laboratory for Physical Sciences.

\end{document}